\documentclass[12pt]{article}
\usepackage{amssymb,amsmath}
\usepackage{graphicx}
\usepackage{epstopdf}
\allowdisplaybreaks
\begin{document}

\title{\bf Dynamics of Polynomial Chaplygin Gas Warm Inflation}
\author{Abdul Jawad \thanks {abduljawad@ciitlahore.edu.pk, jawadab181@yahoo.com}\\
Department of Mathematics, COMSATS Institute of Information\\
Technology, Lahore-54000, Pakistan.\\
Shahid
Chaudhary \thanks{shahidpeak00735@gmail.com}\\
Department of Mathematics, Sharif College of Engineering\\
and Technology, Lahore-54000, Pakistan.\\
Nelson Videla \thanks {nelson.videla@pucv.cl}\\
Instituto de F\'isica, Pontificia Universidad Cat\'olica de Valpara\'iso.
\\ Avda. Universidad 330, Curauma, Valpara\'iso, Chile.}

\date{}
\maketitle
\begin{abstract}

In the present work, we study the consequences of considering a
recently proposed polynomial inflationary potential in the context
of the generalized, modified, and generalized cosmic Chaplygin gas
models. In addition, we consider dissipative effects by coupling the
inflation field to radiation, i.e., the inflationary dynamics is
studied in the warm inflation scenario. We take into account a
general parametrization of the dissipative coefficient $\Gamma$ for
describing the decay of the inflaton field into radiation. By
studying the background and perturbative dynamics in the weak and
strong dissipative regimes of warm inflation separately for the
positive and negative quadratic and quartic potentials, we obtain
expressions for the most relevant inflationary observables as the
scalar power spectrum, the scalar spectral, and the tensor-to-scalar
ratio. We construct the trajectories in the $n_s-r$ plane for
several expressions of the dissipative coefficient and compare with
the two-dimensional marginalized contours for ($n_s,r$) from the
latest Planck data. We find that our results are in agreement with
WMAP9 and Planck 2015 data.

\end{abstract}

\textbf{Keywords:} Warm inflation; Chaplygin gas models; Generalized
dissipative regime; Quadratic and quartic potentials; Inflationary
scenario.

\section{Introduction}

The anisotropies observed in the cosmic microwave background (CMB) are
compatible with an adiabatic primordial scalar
perturbation which is nearly Gaussian with a nearly scale-invariant
power spectrum \cite{1}. Cosmic inflation is the most successful
theoretical framework in describing the very early stages of
universe and also solves some shortcomings of the hot big-bang model, such as horizon, flatness and monopole \cite{2} problems. However, the key feature of
inflation is that it may be able to generate a causal mechanism to explain
the large scale structure (LSS) of the universe \cite{3} and  the  origin  of  the  anisotropies observed in the CMB, since  primordial  density  perturbations  may
be sourced from quantum fluctuations of the inflaton scalar field  during  the  inflationary  expansion. The  standard  cold
inflation  scenario \cite{21}-\cite{24} is  divided  into  two  regimes:  the  slow-roll  and  reheating  phases.  In  the  slow-roll  period  the  universe  undergoes  an  accelerated  expansion  as the potential energy of the inflaton field dominates over
its kinetic energy and  all  interactions  of  the  inflaton scalar  field  with  other  field  degrees
of freedom are typically neglected. Subsequently, a reheating period is invoked to end the brief acceleration. During reheating, kinetic energy of the inflaton field becomes comparable to its potential energy and by transferring its energy to massless particles,
it oscillates around the minimum of the potential. After
reheating, the universe is filled with relativistic particles and then the universe enters in the radiation big-bang epoch.

Alternatively, there is another dynamical mechanism for obtain a successful slow-roll inflation, i.e., the warm inflation scenario \cite{25}-\cite{BasteroGil:2009ec}. As opposed to standard cold inflation, warm inflation has the essential feature that a reheating  phase  is  avoided  at  the  end  of  the  accelerated  expansion due to the decay of the inflaton into radiation and particles during the slow-roll phase. However, the key difference
is the origin of the density fluctuations. In the warm inflation scenario, a thermalized radiation component  is  present  at temperature $T$, which $T>H$, where $H$ is the Hubble rate. In this way, the  inflaton  fluctuations $\delta \phi$
are predominantly thermal instead quantum \cite{25}-\cite{BasteroGil:2009ec}.

Regarding standard cold inflation, Linde \cite{a13Am} introduced the concept of chaotic inflation in
order to interpret the initial conditions for scalar field driving
inflation which may help in solving the persisting problems of the old
inflation models. In this model, the inflaton potential was chosen to be
quadratic or quartic form, i.e. $\frac{m^2}{2}\phi^2$ or $\frac{\lambda}{4}\phi^4$, terms that are always present in the scalar potential of the Higgs sector in all
renormalizable gauge field theories \cite{Pich:2007vu} in which the gauge
symmetry is spontaneously broken via the Englert-Brout-Higgs mechanism \cite{Higgs:1964pj,Englert:1964et}. Such models are interesting for
their simplicity, and has become one of the most favored,
because they predict a significant amount of tensor perturbations due to the inflaton field gets across the trans-Planckian
distance during inflation \cite{Lyth:1996im}.

After introducing chaotic inflation, several works
have been done in this direction. Herrera \cite{a2aAm} discussed the
warm inflation by assuming the chaotic potential in loop quantum
cosmology and found consistency of results with observational data.
The warm inflation was also investigated by del Campo and Herrera
\cite{9sadAm} driving by a scalar field with canonical kinetic term and a power-law dependence in the inflaton field for the dissipative
coefficient, i.e., $\Gamma\propto \phi^{n}$, in the generalized
Chaplygin gas (GCG) scenario.
Further, Setare and Kamali investigated warm inflation driving by
a tachyonic field and assumed the scale factor evolves according to intermediate \cite{2tsadAm} and logamediate \cite{1tsadAm}
models. On the other hand, Bastero-Gill et al. \cite{6sadAm} obtained analytic expressions
for the dissipative coefficient in supersymmetric (SUSY) models and
found that their results provide a realization of
warm inflation in SUSY field theories. After, Bastero-Gill et al.
\cite{5sadAm} have also explored chaotic inflation by assuming the
quartic potential. On the other hand, Herrera et al. studied
intermediate inflation in the context of GCG using standard and
tachyon scalar field models \cite{9tsadAm}. More recently, in Ref.\cite{Jawad:2017gwa}, it was studied the dynamics of warm inflation in a modified Chaplygin gas (MCG).

Panotopoulos and Videla \cite{i17Am} discussed the warm inflation by
assuming the quartic potential and an inflaton decay rate proportional to
temperature and found that their results are in agreement with the latest Planck data, obtaining a lower value for the tensor-to-scalar ratio compared to the cold inflation scenario. Going further, several authors
have investigated the warm inflation scenario in various alternative/modified
theories of gravity \cite{ajAm,bam1Am}. Moreover, a new family of
inflation models is being developed named as shaft inflation
\cite{shaftAm}. In Ref.\cite{hhh}, the authors have developed inflationary parameters by
considering shaft potential and tachyon scalar field and found that
their results are consistent with current observational data. Recently, Kobayashi and Seto \cite{poly} investigated the polynomial
warm inflation and reported that their results are consistence with
BICEP2 and Planck data.

The main goal of the present paper is to study the consequences of considering a
polynomial inflationary potential in the context of the generalized, modified, and generalized
cosmic Chaplygin gas models. In addition, we consider dissipative effects by coupling
the inflation field to radiation, i.e., the inflationary dynamics will be studied in
the warm inflation scenario. We take into account a general parametrization of
the dissipative coefficient $\Gamma$ for describing the decay of the inflaton field into
radiation. The outline of the paper is as
follows: In the next section, we provide the basic set of equations
 describing the warm inflationary scenario. In section \textbf{3}, \textbf{4} and
\textbf{5}, we obtain the several inflationary observables in view of GCG,
modified Chaplygin gas (MCG), and generalized cosmic Chaplygin gas
(GCCG) models for positive/negative quadratic and quartic
potentials. In section \textbf{6}, we summarize our findings and present our conclusions.

\section{Basics of warm inflation scenario}

\subsection{Background evolution}

The Friedmann equation for flat FRW universe in the presence of
standard scalar field an a radiation fluid takes the following form
\begin{equation}\label{1}
H^2=\frac{1}{3M_p{}^2}\left(\rho _{\phi}+\rho_{\gamma }\right),
\end{equation}
where $M_p=\frac{1}{\sqrt{8\text{$\pi $G}}}$ is the reduced Planck
mass. The energy density of standard scalar field can
be defined as $\rho_{\phi}=\dot{\phi}^2/2 +V(\phi)$ where dots
represent derivative with respect to cosmic time. On the other hand, $\rho_{\gamma }$
corresponds to the energy density of radiation fluid. The
corresponding conservation equations for both standard scalar field and radiations turn out to be
\begin{equation}\label{2}
\dot{\rho}_{\phi}+3H(\rho_{\phi}+p_{\phi})=-\Gamma\dot{\phi}^2,\quad
\dot{\rho}_{\gamma}+4H(\rho_{\gamma})=\Gamma\dot{\phi}^2,
\end{equation}
where $p _{\phi}$ denotes the pressure of standard scalar field, given by $p_{\phi}=
\dot{\phi}^2/2 -V(\phi)$. By replacing the expressions for energy densities for
the scalar field as well as radiation in the first conservation equation, we get
\begin{equation}\label{3}
\ddot{\phi}+(3H+\Gamma)\dot{\phi}+V'=0,
\end{equation}
where the primes denote derivatives with respect to $\phi$. On the other hand,
$\Gamma\dot{\phi}$ denotes the interaction term between the scalar field and radiation, whereas $\Gamma$ is the responsible for the decay of the
scalar field into radiation. This inflaton decay rate may depend on scalar
field and temperature of thermal bath, or both quantities, or even it can be a constant.
During warm inflation, the production of radiation is quasi-stable,
i.e., $\dot{\rho}_{\gamma}\ll4H\rho_{\gamma}$ and
$\dot{\rho}_{\gamma}\ll\Gamma\dot{\phi}^2$ \cite{3,5}-\cite{9}. This
implies that energy density related to scalar field dominates over the
energy density of radiation field and hence, the equations of motion
under slow-roll approximation turn out to be
\begin{equation}\label{4a}
3H(1+R)\dot{\phi}\simeq-V',\quad
4H\rho_{\gamma}\simeq\Gamma\dot{\phi}^2,
\end{equation}
where $R=\frac{\Gamma}{3H}$ characterizes two dissipative regimes
such as weak ($R\ll1$) and strong ($R\gg1$). A general parametrization
of the inflaton decay rate is given by
\begin{equation}
\Gamma=c\frac{T^n}{\phi^{n-1}},\label{gamma}
\end{equation} where
$c$ is a constant  parameter and $n$ is an integer \cite{15,16}. Several expressions for the dissipative coefficient, corresponding to
different values of $n$, have been studied in the literatue \cite{28,PRD}. On the other hand, the energy density of the radiation field can be written
as $\rho_{\gamma}=C_{*}T^4$, with $C_{*}=\pi^2 g_{*}/30$ and $g_{*}$
represents the number of relativistic degrees of freedom. In minimal
SUSY standard model, $g_{*}=228.75$ and $C_{*}\simeq 70$ . By using
Eq.(\ref{4a}) and $\rho_{\gamma}\propto T^4$, we may find an expression for the temperature of thermal bath which is given by
\begin{equation}\label{5}
T=\bigg(\frac{\Gamma V'^2 }{6^2C_{*}H^3(1+R)^2}\bigg)^\frac{1}{4}.
\end{equation}
The set of slow-roll parameters for warm inflation is given by \cite{5}
\begin{equation}\label{slow}
\epsilon=\frac{-\dot{H}}{H^2},\quad
\eta=\frac{-\ddot{H}}{H\dot{H}},\quad
\beta=\frac{-1}{H}\frac{d}{dt}(\ln\Gamma).
\end{equation}

The number of $e$-folds is defined as
\begin{equation}\label{6}
N=\int_{t_{*}}^{t_{end}} Hdt.
\end{equation}
where $t_{*}$ and $t_{end}$ denote the moment when the cosmological scales crosses the
Hubble-radius and the end of inflation, respectively.

\subsection{Cosmological perturbations}

In the warm inflation scenario, a thermalized radiation component is present with
$T>H$, then the inflaton fluctuations $\delta \phi$
are predominantly thermal instead quantum.  In this way,
following Ref.\cite{BasteroGil:2009ec}, the amplitude of the power spectrum of the curvature perturbation is given by
\begin{equation}\label{7}
\mathcal{P}_{\mathcal{R}}\simeq \left(\frac{H}{2\pi}\right)\left(\frac{3H^2}{V^{\prime}}\right)(1+R)^{5/4}\left(\frac{T}{H}\right)^{1/2},
\end{equation}
where  the  normalization  has  been  chosen  in  order  to  recover  the  standard  cold  inflation result when $R\,\rightarrow\,0$ and $T\simeq H$.

By the other hand, the scalar spectral index $n_s$,
to leading order in the slow-roll approximation, becomes
\begin{equation}\label{8}
n_{s}=1+\frac{dP_{R}}{dlnk}\simeq 1-\frac{9\epsilon}{4}+\frac{3\eta}{2}-\frac{9\beta}{4},
\end{equation}
while for the tensor-to-scalar ratio, we have that \cite{BasteroGil:2009ec}
\begin{equation}\label{9}
r\simeq \left(\frac{H}{T}\right)\frac{16 \epsilon}{(1+R)^{5/2}}.
\end{equation}

When a specific form of the scalar potential and the dissipative coefficient Γ are considered,  it is possible to study the background evolution under the slow-roll regime and the primordial perturbations in order to test the viability of warm inflation. In the following we will study a polynomial potential,
which has quadratic and quartic powers
of the inflaton scalar field. A generalized expressions for the polynomial
potential is proposed in \cite{poly}, given by
\begin{equation}\label{pot1}
V=t_{1}+t_{2}\phi^{2} +t_{4}\phi^{4}.
\end{equation}
Since it is not easy to deal with several parameters, for convenience, we
consider terms up to $\phi^{4}$ which might be motivated by the
renormalizability for this potential in quantum field theory. Thus, we consider
two kind of polynomial potentials:
\begin{itemize}
\item \underline{Negative quadratic and quartic potential}

\begin{equation}\label{pot11}
V=s-\frac {1}{2}\sigma^2\phi^2+\frac{\lambda_{*}}{4}\phi^4, \quad
V'=-\sigma^2\phi+\lambda_{*}\phi^3
\end{equation}

\item \underline{Positive quadratic and quartic
potential}

\begin{equation}\label{pot111}
V=\frac{1}{2}\sigma^2\phi^2+\frac{\lambda_{*}}{4}\phi^4, \quad
V'=\sigma^2\phi+\lambda_{*}\phi^3
\end{equation}

\end{itemize}
where $s,~\lambda_{*}$ and $\sigma$ are arbitrary constants. In the
following sections, we illustrate the inflationary parameters for
above mentioned scenario in the presence of three Chaplygin gas
models (GCG, MCG and GCCG) by assuming weak and strong dissipative
regimes with the inflaton decay rate given by the generalized expression (\ref{gamma}).

\section{Generalized Chaplygin Gas Model}

It is believed that the universe undergoes an accelerated expansion
of the universe and an exotic component having a negative pressure,
usually known as dark energy (DE), is responsible for this
expansion. Several models have been already proposed to be DE
candidates, such as cosmological constant \cite{de1}, quintessence
\cite{de2}-\cite{de4}, k-essence \cite{de5}-\cite{de7}, tachyon
\cite{de8}-\cite{de10}, phantom \cite{de11}-\cite{de13}, Chaplygin
gas \cite{de14}, holographic DE \cite{Li:2004rb}, among others in
order to modify the matter sector of the gravitational action.
Despite the plenty of models, the nature of the dark sector of the
universe, i.e. DE and dark matter, is still unknown. There exists
another way of understanding the observed universe in which dark
matter and DE are described by a single unified component.
Particularly, CG contains the unification of DE and dark matter and
behaves as a pressureless matter at the early times and like a
cosmological constant at late times \cite{de14}. It is noted that CG
describes the universe in agreement with current observations of
cosmic acceleration. The GCG is the extended form of CG and its
equation of state (EoS) is \cite{de14}
\begin{equation}\label{cg1}
p_{gcg}=-\frac{C_{1}}{\rho_{gcg}^\alpha}
\end{equation}
where $p_{gcg}$ and $\rho_{gcg}$ represent the pressure and energy
density of GCG model, respectively, with $0<\alpha\leq1$ and $C_{1}$
is the positive constant \cite{de14}. Also, $\rho_{gcg}$ can be
obtained through conservation equation as follows
\begin{equation}\label{10}
\rho_{gcg}=\bigg(C_{1}+\frac{C_{2}}{a^{3(1+\alpha)}}\bigg)^\frac{1}{1+\alpha},
\end{equation}
here $C_{2}$ is positive constant after integration. In this way, the term proportional to $a^{-3}$ is identified as the energy  density  of  matter $\rho_m$.

In order to obtain inflation in the Chaplygin-like gas scenarios studied in the present work, we follow Ref.\cite{Bertolami:2006zg}. The energy density of matter $\rho_m$ is identified with  the  contribution  of  the  energy  density associated to the standard scalar field $\rho_{\phi}$ through an extrapolation of
Eq.(\ref{10}), yielding
\begin{equation}\label{12c}
\rho_{gcg}=(C_{1}+\rho^{1+\alpha}_{m})^\frac{1}{1+\alpha}\quad
\longrightarrow (C_{1}+\rho^{1+\alpha}_{\phi})^\frac{1}{1+\alpha}.
\end{equation}

In this sense, we will not consider Eq.(\ref{12c}) as a consequence of Eq.(\ref{10}), but a non-covariant modification of gravity instead, resulting in a modified Friedmann equation, as it was pointed up in Ref.\cite{Barreiro:2004bd}.

In this scenario,  we consider a spatially flat universe which contains a self-interacting inflation field $\phi$ and a radiation field, then the Friedmann equation (\ref{1}) turns out to be
\begin{equation}\label{11}
H^2=\frac{1}{3M^2_p}\bigg[(C_{1}+\rho_\phi^{1+\alpha})^\frac{1}{1+\alpha}+\rho_{\gamma}\bigg].
\end{equation}

We stress that Friedmann equation (\ref{11}) comes from a non-covariant modification of gravity. However,  as  it  was  pointed  up  in Ref.\cite{Bertolami:2006zg},  it  may  assumed  that  the  effect  giving  rise  to
Eq. (\ref{11}) preserves diffeomorphism invariance in (3+1) dimensions, whence total stress-energy conservation follows.

During inflation,
the energy density of the scalar field dominates the energy density
of the radiation field, i.e., $\rho_{\phi\gg}\rho_{\gamma}$ which
leads to $\rho_{\phi}\sim V$. Here we take $\alpha=1$ for the sake
of simplicity. Then the Friedmann equation (\ref{11}) becomes

\begin{equation}\label{13}
H^2=\frac{1}{3M^2_p}\sqrt{C_{1}+\rho^2_{\phi}}\sim\frac{1}{3M^2_p}\sqrt{C_{1}+V^2}.
\end{equation}
Further, we will construct the inflationary parameters under two
cases of dissipative coefficient that is weak dissipative regime
($R\leq1$) and strong dissipative regime ($R\geq1$) for the present
case of GCG.

\subsection{Weak Dissipative Regime}

For this case, temperature turns out to be
$T=(\frac{cV'^2}{6^2C_{*}H^3\phi^{n-1}})^\frac{1}{4-n}$ in the
presence of generalized dissipative coefficient
$\Gamma=c\frac{T^n}{\phi^{n-1}}$. With this temperature and weak
dissipative regime condition, the slow roll parameters (\ref{slow})
in terms of potential ($V$) can be written as
\begin{eqnarray}\nonumber
\epsilon&=&\frac{M^2_pVV'^2}{2(C_{1}+V^2)^\frac{3}{2}},\quad
\eta=\frac{M^2_p}{(C_{1}+V^2)^\frac{1}{2}}\bigg(V''+\frac{V'^2}{V}-\frac{3VV'^2}{2(C_{1}+V^2)}\bigg),\\\nonumber
\beta&=&M_p^2\bigg(2(C_{1}+V^2)(2nV'' -n V'(n-1)\phi
^{-1}-V'(4-n)(n-1)\\\nonumber&\times&\phi ^{-1})-3n
V'^2V\bigg)\bigg({2(C_{1}+V^2)^{\frac{3}{2}}(4-n)}\bigg)^{-1}.
\end{eqnarray}
With the help of Eq.(\ref{6}), the number of e-folds become
\begin{equation}\nonumber
N=\frac{1}{M^2_p}\int_{\phi_{end}}^{\phi_{e}}\frac{\sqrt{C_{1}+V^2}}{V'}d\phi.
\end{equation}
Equations (\ref{7})-(\ref{9}) provide the power spectrum, scalar
spectral index and tensor-to-scalar ratio in terms of potential
($V$) as follow
\begin{eqnarray}\nonumber
\mathcal{P}_{\mathcal{R}}&=&\left(\frac{\pi
}{4}\right)^{\frac{1}{2}}\frac{9\left(C_1+V^2\right)^{\frac{3(5-2n)}{4(4-n)}}
c^{\frac{3}{4-n}}}{3^{\frac{3(5-2n)}{2(4-n)}}
M_p^{\frac{3(5-2n)}{4-n}}V'^{\frac{6-3n}{4-n}}\phi
^{\frac{(n-1)(4-n)+(n-1)(n+2)}{2(4-n)}}6^{\frac{n+2}{4-n}}C_{*}^{\frac{n+2}{2(4-n)}}},\\\nonumber
n_s-1&=&\frac{3M_p^2}{2(C_{1}+V^2)^{\frac{1}{2}}}\bigg(\frac{-9V
V'^2}{4(C_{1}+V^2)}-\frac{3}{2}\bigg(2(C_{1}+V^2)\bigg(2n
V''\\\nonumber&-&n V'(n-1)\phi ^{-1}-V'(4-n)(n-1)\phi ^{-1}\bigg)-3n
V'^2V\bigg)\\\nonumber&\times&\bigg({2(C_{1}+V^2)(4-n)}\bigg)^{-1}+V''+\frac{V'^2}{V}\bigg),\\\label{16}
r&=&\frac{32\text{G V}^{ \prime ^{\frac{6-3n}{4-n}}}\phi
^{\frac{(n-1)(4-n)+(n-1)(n+2)}{2(4-n)}}6^{\frac{n+2}{4-n}}C_{*}^{\frac{n+2}{2(4-n)}}
3^{\frac{7-4n}{2(4-n)}}M_p^{\frac{7-4n}{4-n}}}{9c^{\frac{3}{4-n}}\pi
^{\frac{3}{2}}\left(C_{1}+V^2\right)^{\frac{7-4n}{4(4-n)}}}.
\end{eqnarray}
\underline{For positive quadratic and quartic potential:} The above
expressions of $r$ and $n_{s}$ in terms of scalar field lead to
\begin{eqnarray}\nonumber
r&=&32G\bigg(\sigma^2\phi +\lambda_{*} \phi
^3\bigg)^{^{\frac{6-3n}{4-n}}}\phi
^{\frac{(n-1)(4-n)+(n-1)(n+2)}{2(4-n)}}6^{\frac{n+2}{4-n}}C_{*}^{\frac{n+2}{2(4-n)}}
3^{\frac{7-4n}{2(4-n)}}\\\nonumber&\times&M_p^{\frac{7-4n}{4-n}}\bigg(9c^{\frac{3}{4-n}}\pi
^{\frac{3}{2}}(C_{1}+(\frac{\sigma^2\phi ^2}{2}+\frac{\lambda_{*}
\phi ^4}{4})^2)^{\frac{7-4n}{4(4-n)}}\bigg)^{-1},\\\nonumber
n_s-1&=&\frac{3M_p^2}{2(C_{1}+(\frac{\sigma^2\phi
^2}{2}+\frac{\lambda_{*} \phi
^4}{4})^2)^{\frac{1}{2}}}\bigg(-9(\frac{\sigma^2\phi
^2}{2}+\frac{\lambda_{*} \phi ^4}{4})(\sigma^2\phi
\\\nonumber&+&\lambda_{*} \phi
^3)^2\bigg({4(C_{1}+(\frac{\sigma^2\phi ^2}{2}+\frac{\lambda_{*}
\phi ^4}{4})^2)}\bigg)^{-1}-\frac{3}{2}\bigg(
2(C_{1}+(\frac{\sigma^2\phi ^2}{2}\\\nonumber&+&\frac{\lambda_{*}
\phi ^4}{4})^2)(2n (\sigma^2+3\lambda_{*} \phi ^2)-n(\sigma^2\phi
+\lambda_{*} \phi ^3)(n-1)\phi ^{-1}\\\nonumber&-&(\sigma^2\phi
+\lambda_{*} \phi ^3)(4-n)(n-1)\phi ^{-1})-3n(\sigma^2\phi
+\lambda_{*} \phi ^3)^2\\\nonumber&\times&(\frac{\sigma^2\phi
^2}{2}+\frac{\lambda_{*} \phi ^4}{4})(2(2(C_{1}+(\frac{\sigma^2\phi
^2}{2}+\frac{\lambda_{*} \phi
^4}{4})^2))(4\\\nonumber&-&n))^{-1}\bigg)+(\sigma^2+3\lambda_{*}
\phi ^2)+\frac{(\sigma^2\phi +\lambda_{*} \phi
^3)^2}{(\frac{\sigma^2\phi ^2}{2}+\frac{\lambda_{*} \phi
^4}{4})}\bigg).
\end{eqnarray}
\underline{For negative quadratic and quartic potential:} The
expressions of $r$ and $n_{s}$ for negative potential turn out to be
\begin{eqnarray}\nonumber
r&=&\frac{32G\left(-\sigma^2\phi +\lambda_{*} \phi
^3\right)^{^{\frac{6-3n}{4-n}}}\phi
^{\frac{(n-1)(4-n)+(n-1)(n+2)}{2(4-n)}}6^{\frac{n+2}{4-n}}C_{*}^{\frac{n+2}{2(4-n)}}
3^{\frac{7-4n}{2(4-n)}}M_p{}^{\frac{7-4n}{4-n}}}{9c^{\frac{3}{4-n}}\pi
^{\frac{3}{2}}\left(C_{1}+\left(s-\frac{\sigma^2\phi
^2}{2}+\frac{\lambda_{*} \phi
^4}{4}\right)^2\right)^{\frac{7-4n}{4(4-n)}}},\\\nonumber
n_s&=&1+\frac{3M_p^2}{2(C_{1}+(s-\frac{\sigma^2\phi
^2}{2}+\frac{\lambda_{*} \phi
^4}{4})^2)^{\frac{1}{2}}}\bigg(-9(s-\frac{\sigma^2\phi
^2}{2}+\frac{\lambda_{*} \phi ^4}{4})\\\nonumber&\times&
(-\sigma^2\phi +\lambda_{*} \phi
^3)^2({4(C_{1}+(s-\frac{\sigma^2\phi ^2}{2}+\frac{\lambda_{*} \phi
^4}{4})^2)})^{-1}-\frac{3}{2}\\\nonumber&\times&\bigg(2(C_{1}+(s-\frac{\sigma^2\phi
^2}{2}+\frac{\lambda_{*} \phi ^4}{4})^2)(2n(-\sigma^2+3\lambda_{*}
\phi ^2)-n\\\nonumber&\times&(-\sigma^2\phi +\lambda_{*} \phi
^3)(n-1)\phi ^{-1}-(-\sigma^2\phi +\lambda_{*} \phi
^3)(4-n)\\\nonumber&\times&(n-1)\phi ^{-1})-3n(-\sigma^2\phi
+\lambda_{*} \phi ^3)^2(s-\frac{\sigma^2\phi
^2}{2}+\frac{\lambda_{*} \phi
^4}{4})\\\nonumber&\times&({2(2(C_{1}+(s-\frac{\sigma^2\phi
^2}{2}+\frac{\lambda_{*} \phi
^4}{4})^2))(4-n))^{-1}}\bigg)+(-\sigma^2\\\nonumber&+&3\lambda_{*}
\phi ^2)+\frac{(-\sigma^2\phi +\lambda_{*} \phi
^3)^2}{(s-\frac{\sigma^2\phi ^2}{2}+\frac{\lambda_{*} \phi
^4}{4})}\bigg).
\end{eqnarray}

\subsection{Strong Dissipative Regime}

In this case, the temperature becomes
$T=(\frac{V'^2\phi^{n-1}}{4cc_{*}H})^\frac{1}{n+4}$. The slow-roll
parameters lead to
\begin{eqnarray}\nonumber
\epsilon&=&\frac{M^2_pVV'^2}{2R(C_{1}+V^2)^\frac{3}{2}},\quad
\eta=\frac{M^2_p}{R(C_{1}+V^2)^\frac{1}{2}}\bigg(V''+\frac{V'^2}{V}-\frac{3VV'^2}{2(C_{1}+V^2)}\bigg),\quad\\\nonumber
\beta&=&\frac{nM^2_P}{(n+4)R}(\frac{4V''(C_{1}+V^2)-VV'^2}{2(C_{1}+V^2)^\frac{3}{2}}).
\end{eqnarray}
The power spectrum, scalar spectral index and tenor-to-scalar ratio
turn out to be
\begin{eqnarray}\nonumber
\mathcal{P}_{\mathcal{R}}&=&\left(\frac{\pi
}{4}\right)^{\frac{1}{2}}\frac{c^{\frac{23-5n}{10}}\left(C_{1}+V^2\right)^{\frac{23-5n}{40}}}
{V'^{\frac{8-5n}{5}}(4c_{*})^{\frac{5n+2}{10}}
M_p{}^{\frac{23-5n}{10}}3^{\frac{23-5n}{20}}},\\\nonumber n_s-1&=&
\bigg(\frac{3(4c_{*})^{\frac{1}{5}}}{c^{\frac{4}{5}}V'^{\frac{2}{5}}}\bigg)\frac{(3M_p^2)
^{\frac{2}{5}}}{2(C_{1}+V^2)^{\frac{1}{5}}}\bigg(V''+\frac{V^{\prime
^2}}{V}-\frac{9V V'^2}{4(C_{1}+V^2)}
\\\nonumber&-&\frac{3(n)}{2(n+4)}\bigg(\frac{4V''(C_{1}+V^2)-V V^{\prime
^2}}{2(C_{1}+V^2)}\bigg)\bigg)\\\label{19} r&=&\frac{32G V^{\prime
^{\frac{8-5n}{5}}}\phi
^{\frac{5(n-1)}{2}}(4c_{*})^{\frac{5n+2}{10}}\left(C_{1}+V^2\right)^{\frac{5n-3}{40}}}{\pi
^{\frac{3}{2}}c^{\frac{23-5n}{10}}3^{\frac{5n-3}{20}}M_p^{\frac{5n-3}{10}}}.
\end{eqnarray}
\underline{For positive quadratic and quartic potential:} The scalar
spectral index and tensor-to-scalar ratio in case of strong
dissipative regime turns out to be
\begin{eqnarray}\nonumber
r&=&\frac{32G \left(\sigma^2\phi +\lambda_{*} \phi
^3\right)^{^{\frac{8-5n}{5}}}\phi
^{\frac{5(n-1)}{2}}(4c_{*})^{\frac{5n+2}{10}}\left(C_{1}+\left(\frac{\sigma^2\phi
^2}{2}+\frac{\lambda_{*} \phi
^4}{4}\right)^2\right)^{\frac{5n-3}{40}}}{\pi
^{\frac{3}{2}}b^{\frac{23-5n}{10}}3^{\frac{5n-3}{20}}M_p^{\frac{5n-3}{10}}}\\\nonumber
n_s-1&=&
\bigg(\frac{3(4c_{*})^{\frac{1}{5}}}{c^{\frac{4}{5}}(\sigma^2\phi
+\lambda_{*} \phi
^3)^{\frac{2}{5}}}\bigg)\frac{(3M_p^2)^{\frac{2}{5}}}{2(C_{1}+(\frac{\sigma^2\phi
^2}{2}+\frac{\lambda_{*} \phi
^4}{4})^2)^{\frac{1}{5}}}\bigg((\sigma^2\\\nonumber&+&3\lambda_{*}
\phi ^2)+\frac{(\sigma^2\phi +\lambda_{*} \phi
^3)^{^2}}{(\frac{\sigma^2\phi ^2}{2}+\frac{\lambda_{*} \phi
^4}{4})}-\frac{9(\frac{\sigma^2\phi ^2}{2}+\frac{\lambda_{*} \phi
^4}{4})(\sigma^2\phi +\lambda_{*} \phi
^3)^2}{4(C_{1}+(\frac{\sigma^2\phi ^2}{2}+\frac{\lambda_{*} \phi
^4}{4})^2)}\\\nonumber&-&\frac{3(n)}{2(n+4)}\bigg(4(\sigma^2+3\lambda_{*}
\phi ^2)(C_{1}+(\frac{\sigma^2\phi ^2}{2}+\frac{\lambda_{*} \phi
^4}{4})^2)\\\nonumber&-&(\frac{\sigma^2\phi ^2}{2}+\frac{\lambda_{*}
\phi ^4}{4})(\sigma^2\phi +\lambda_{*} \phi
^3)^{^2}(2(C_{1}+(\frac{\sigma^2\phi
^2}{2}\\\label{12}&+&\frac{\lambda_{*} \phi
^4}{4})^2))^{-1}\bigg)\bigg).
\end{eqnarray}

\begin{figure}
\includegraphics[width=.45\linewidth]{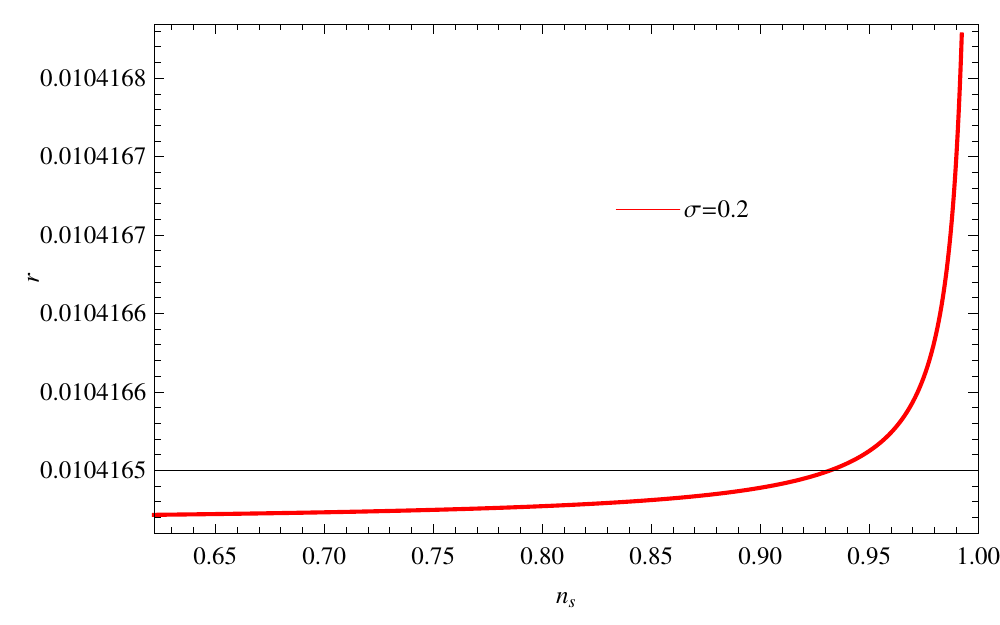}
\includegraphics[width=.44\linewidth]{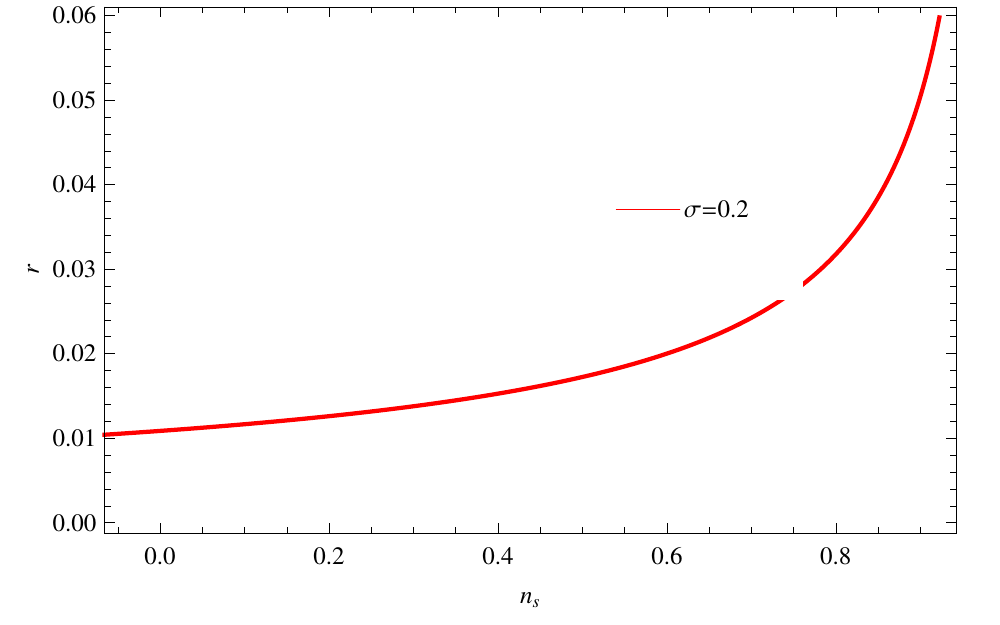}
\caption{Plot of $r$
versus $n_{s}$ for GCG model in weak (left panel) and strong (right
panel) dissipative regimes for positive potential with
$n=1$.}
\end{figure}
\begin{figure}
\includegraphics[width=.45\linewidth]{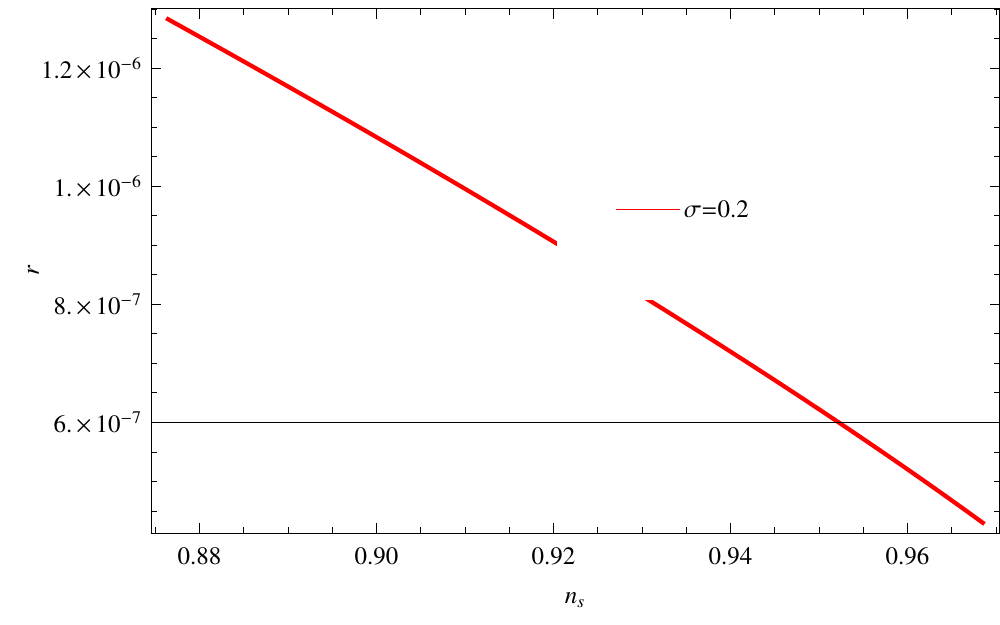}
\includegraphics[width=.45\linewidth]{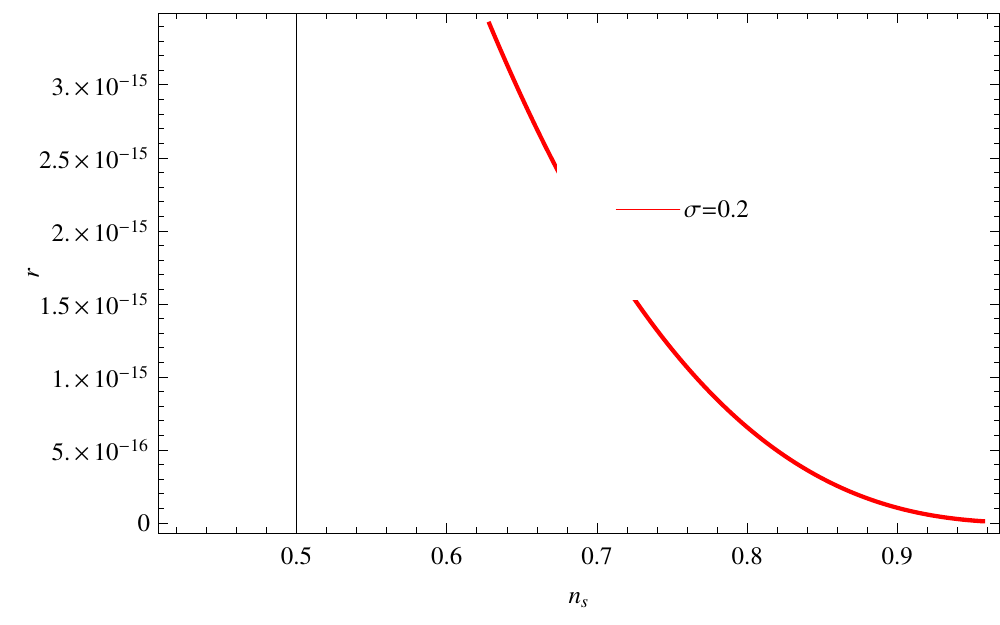}
\caption{Plot of $r$
versus $n_{s}$ for GCG model in weak (left panel) and strong (right
panel) dissipative regimes for positive potential with $n=-1$.}
\end{figure}
\begin{figure}
\includegraphics[width=.45\linewidth]{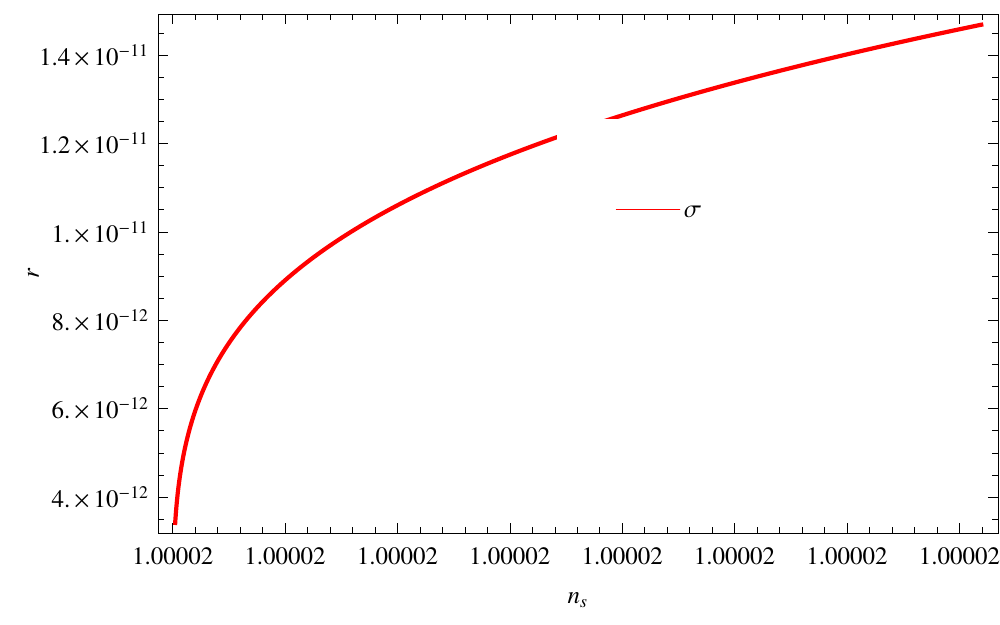}
\includegraphics[width=.45\linewidth]{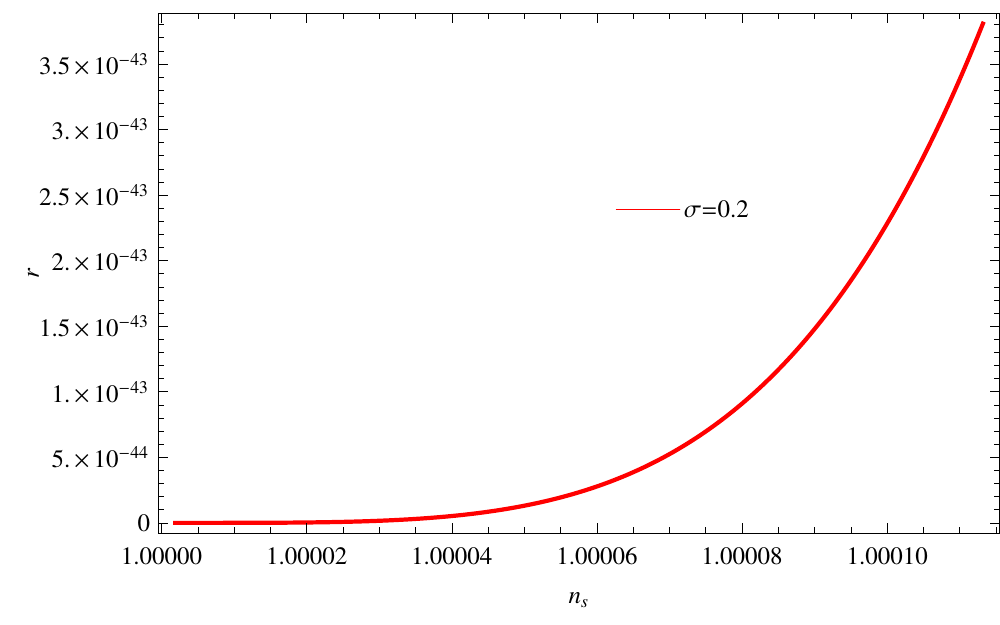}
\caption{Plot of $r$
versus $n_{s}$ for GCG model in weak (left panel) and strong (right
panel) dissipative regimes for positive potential with $n=-2$.}
\end{figure}

By utilizing the value of \underline{negative quadratic and quartic
potential} and its derivative in the expressions (\ref{19}) of the
scalar spectral index and tensor-to-scalar ratio, we obtain
\begin{figure}
\includegraphics[width=.45\linewidth]{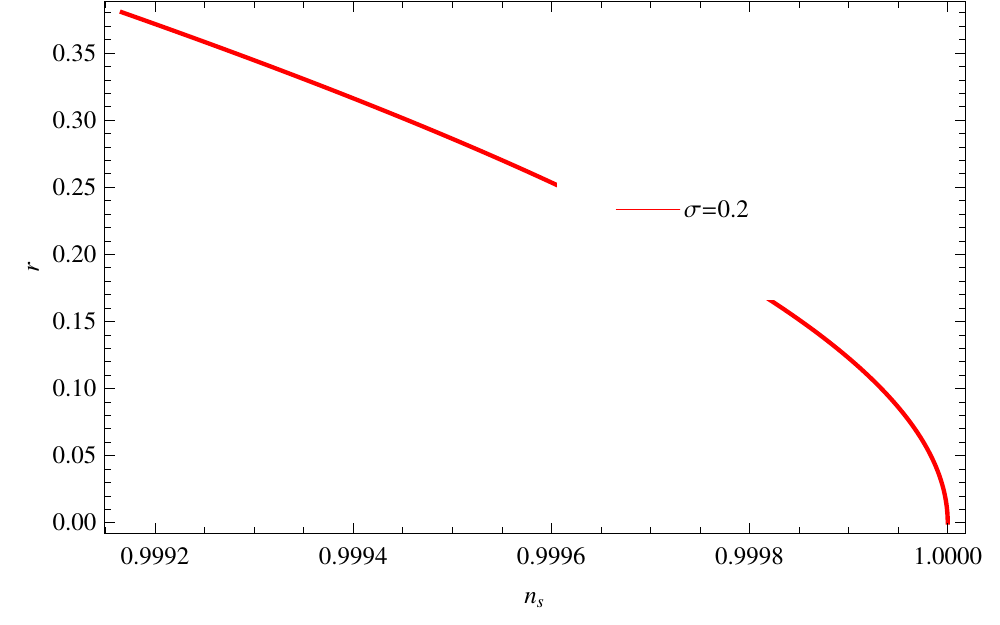}
\includegraphics[width=.45\linewidth]{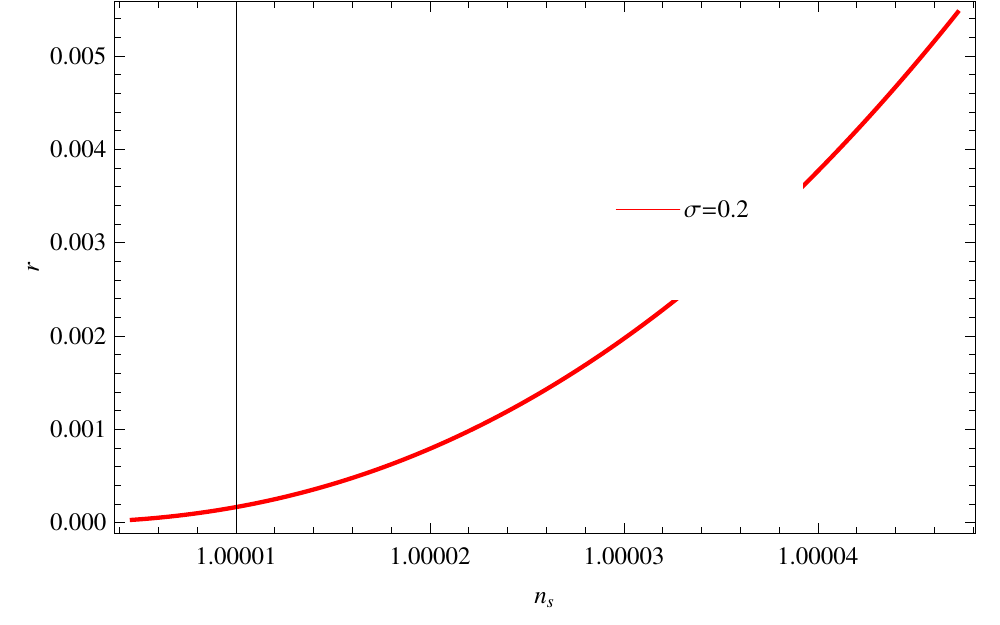}
\caption{Plot of $r$
versus $n_{s}$ for GCG model in weak (left panel) and strong (right
panel) dissipative regimes for negative potential with $n=1$.}
\end{figure}
\begin{figure}
\includegraphics[width=.45\linewidth]{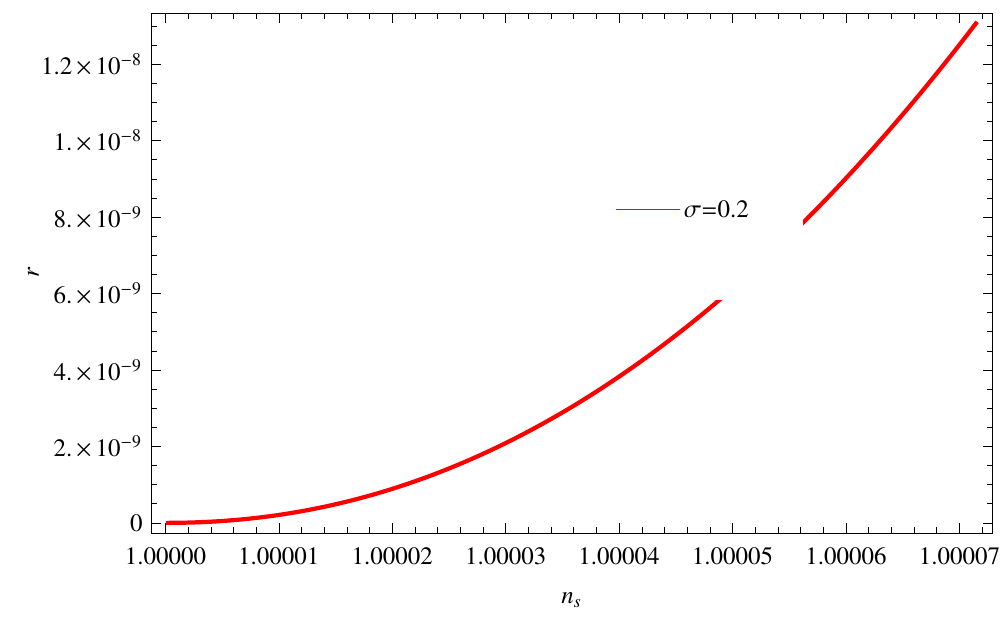}
\includegraphics[width=.45\linewidth]{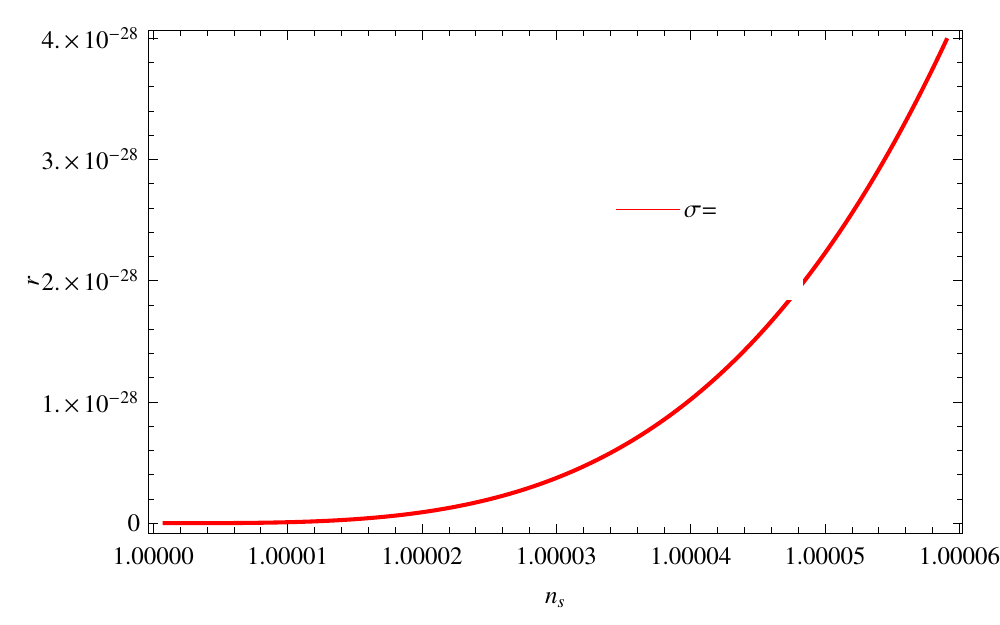}
\caption{Plot of $r$
versus $n_{s}$ for GCG model in weak (left panel) and strong (right
panel) dissipative regimes for negative potential with $n=-1$.}
\end{figure}
\begin{figure}
\includegraphics[width=.45\linewidth]{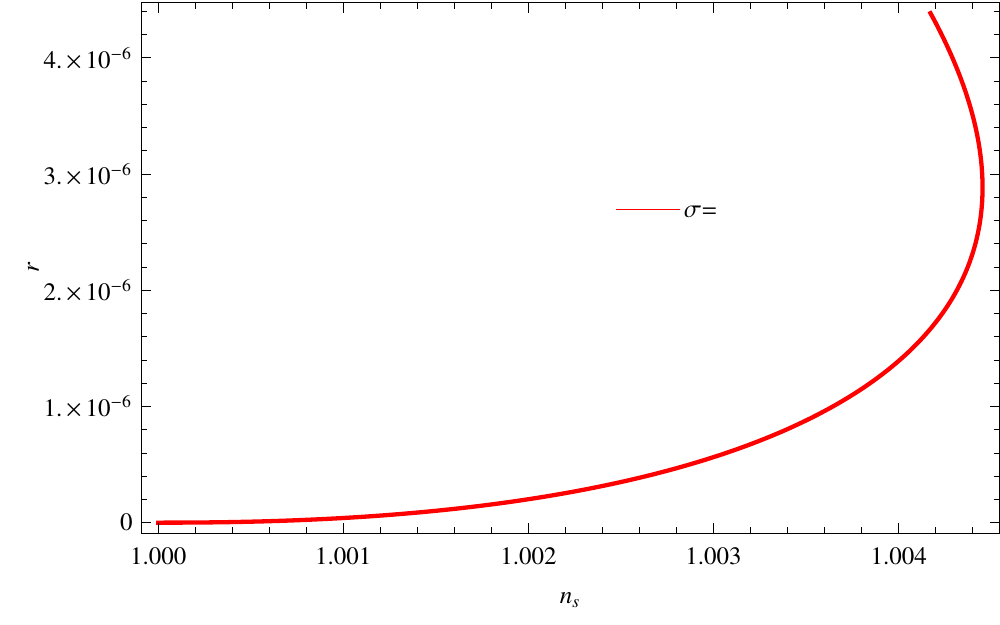}
\includegraphics[width=.45\linewidth]{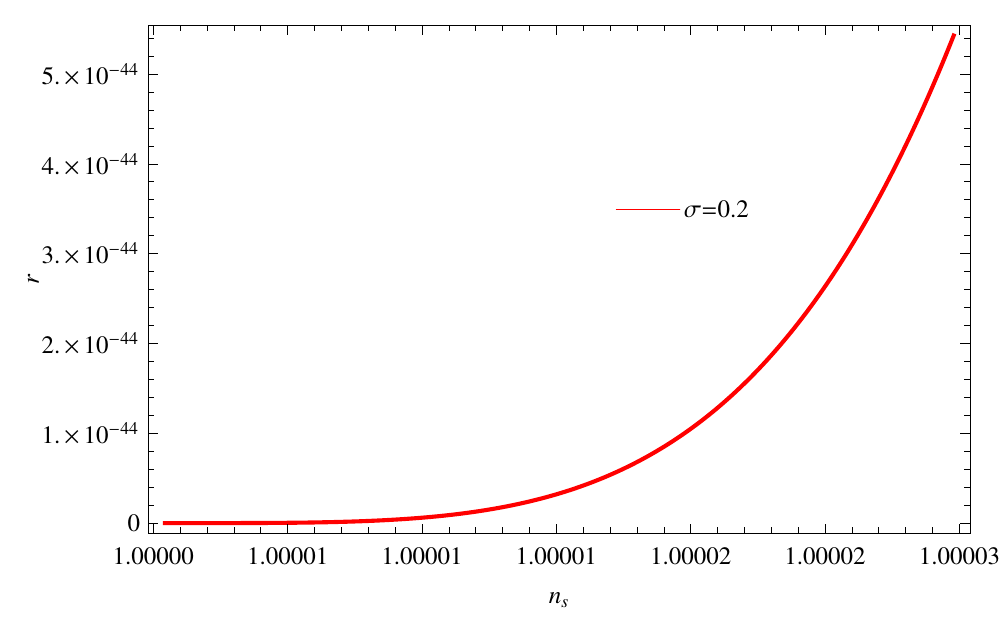}
\caption{Plot of $r$
versus $n_{s}$ for GCG model in weak (left panel) and strong (right
panel) dissipative regimes for negative potential with $n=-2$.}
\end{figure}

\begin{eqnarray}\nonumber
r&=&\frac{32G \left(-\sigma^2\phi +\lambda_{*} \phi
^3\right)^{^{\frac{8-5n}{5}}}\phi
^{\frac{5(n-1)}{2}}(4c_{*})^{\frac{5n+2}{10}}(C_{1}+(\text{s}-\frac{\sigma^2\phi
^2}{2}+\frac{\lambda_{*} \phi ^4}{4})^2)^{\frac{5n-3}{40}}}{\pi
^{\frac{3}{2}}c^{\frac{23-5n}{10}}3^{\frac{5n-3}{20}}M_p^{\frac{5n-3}{10}}}\\\nonumber
n_s&=&1+
\bigg(\frac{3(4c_{*})^{\frac{1}{5}}}{c^{\frac{4}{5}}(-\sigma^2\phi
+\lambda_{*} \phi
^3)^{\frac{2}{5}}}\bigg)\frac{(3M_p^2)^{\frac{2}{5}}}{2(C_{1}+(\text{s}-\frac{\sigma^2\phi
^2}{2}+\frac{\lambda_{*} \phi
^4}{4})^2)^{\frac{1}{5}}}\\\nonumber&\times&\bigg((-\sigma^2+3\lambda_{*}
\phi ^2)+\frac{(-\sigma^2\phi +\lambda_{*} \phi
^3)^{^2}}{(\text{s}-\frac{\sigma^2\phi ^2}{2}+\frac{\lambda_{*} \phi
^4}{4})}-9(\text{s}-\frac{\sigma^2\phi
^2}{2}\\\nonumber&+&\frac{\lambda_{*} \phi ^4}{4})(-\sigma^2\phi
+\lambda_{*} \phi ^3)^2({4(C_{1}+(\text{s}-\frac{\sigma^2\phi
^2}{2}+\frac{\lambda_{*} \phi
^4}{4})^2)})^{-1}\\\nonumber&-&\frac{3(n)}{2(n+4)}\bigg(4(-\sigma^2+3\lambda_{*}
\phi ^2)(C_{1}+(\text{s}-\frac{\sigma^2\phi ^2}{2}+\frac{\lambda_{*}
\phi ^4}{4})^2)\\\nonumber&-&(\text{s}-\frac{\sigma^2\phi
^2}{2}+\frac{\lambda_{*} \phi ^4}{4})(-\sigma^2\phi +\lambda_{*}
\phi ^3)^{^2}(2(C_{1}+(\text{s}-\frac{\sigma^2\phi
^2}{2}\\\label{13}&+&\frac{\lambda_{*} \phi
^4}{4})^2))^{-1}\bigg)\bigg).
\end{eqnarray}
We plot the graphs of $r$ and $n_{s}$ for both weak and strong
dissipative regimes by assuming the positive quadratic and quartic
potential for three different values of $n=1,-1,-2$, respectively as
shown in Figures \textbf{1-3}. The other constant parameters appear
in the model are assumed to be as follows: $M_p=1, \alpha=10^{-13},
C_{1}=10^{-45}, \lambda_{*}=10^{-3}, c=0.3, C_{*}=70$. The
constraints of inflationary parameters such as tensor-to-scalar
ratio and spectral index are displayed in the Table \textbf{1(a)}.
It is found that the results are compatible with WMAP$9$ \cite{i31}
and Planck $2015$ \cite{i32} which are given in Table \textbf{1}.\\\\
\textbf{Table 1:}  WMAP$9$ \cite{i31} and Planck $2015$ \cite{i32}
data for $r$ and $n_s$.
\begin{table}[bht]\centering
\begin{small}
\begin{tabular}{|c|c|c|c|c|}
\hline $r$~~(WMAP$9)$ & $n_{s}$~~(WMAP9)&$r$~~(Planck 2015)&$n_{s}$~~(Planck2015)\\
\hline$<0.13$&$0.972\pm0.013$&$<0.11$&$0.968\pm0.006$\\
\hline
\end{tabular}
\end{small}
\end{table}\\\\
\textbf{Table 1(a):} GCG Weak and Strong Dissipative Regimes with
Positive Quadratic and Quartic Potential.
\begin{table}[bht]\centering
\begin{small}
\begin{tabular}{|c|c|c|c|c|c|c|}
\hline $Sr.No$& $n$& $r(W)$ & $n_{s} (W)$&$r(S)$&$n_{s} (S)$\\
\hline$1$& $1$&$\leq0.0104168$&$0.98_{-0.01}^{+0.01}$&$\leq0.06$&$0.90_{-0.06}^{+0.06}$\\
\hline$2$& $-1$&$\leq1.2\times10^{-6}$&$0.96_{-0.02}^{+0.02}$&$\leq3.0\times10^{-15}$&$0.94_{-0.02}^{+0.02}$\\
\hline$3$&$-2$&$\leq1.4\times10^{-14}$&$
1.00001_{-0.00001}^{+0.00001}$&$\leq3.7\times10^{-43}$&$1.00011_{-0.00001}^{+0.00001}$\\
\hline
\end{tabular}
\end{small}
\end{table}\\\\

Figures \textbf{4-6} show the behavior of $r$ and $n_{s}$ for
negative quadratic and quartic potential in weak and strong
dissipative regimes. The parameters appear in the model attained the
values $M_p=1, s=2$, $\alpha=10^{-5}$, $C_{1}=10^{-45}, c=0.3,
C_{*}=70, \lambda_{*}=10^{-13}, c=10^{-7}, G=0.0398$. The result for
negative quadratic and quartic potential are shown in the Table
\textbf{1(b)}. The values of tensor-to-scalar ratio and spectral
index are consistence with observational data WMAP$9$ \cite{i31} and Planck $2015$ \cite{i32}.\\\\
\textbf{Table 1(b):} GCG Weak and Strong Dissipative Regime with
Negative Quadratic and Quartic Potential.
\begin{table}[bht]\centering
\begin{small}
\begin{tabular}{|c|c|c|c|c|c|c|}
\hline $Sr.No$& $n$& $r(W)$ & $n_{s}(W)$&$r(S)$&$n_{s}(S)$\\
\hline$1$& $1$&$\leq0.38$&$1.0000_{-0.0001}^{+0.0001}$&$\leq0.0055$&$1.0004_{-0.0001}^{+0.0001}$\\
\hline$2$& $-1$&$\leq0.000041$&$1.0031_{-0.0001}^{+0.0001}$&$\leq4.0\times10^{-28}$&$1.000057_{-0.000001}^{+0.000001}$\\
\hline$3$&$-2$&$\leq4.5\times10^{-6}$&$1.0040_{0.0001}^{+0.0001}$&$\leq5.2\times10^{-44}$&$1.00002_{-0.00001}^{+0.00001}$\\
\hline
\end{tabular}
\end{small}
\end{table}\\\\

\section{Modified Chaplygin Gas}

The EoS of MCG is given as \cite{Benaoum:2002zs}
\begin{equation}\label{mcg}
p_{mcg}=\zeta\rho_{mcg}-\frac{\xi}{\rho_{mcg}^\alpha},
\end{equation}
where $p_{mcg}$ and $\rho_{mcg}$ represent the pressure and energy
density of MCG, respectively and $0\leq\alpha\leq1$, while $\zeta$
and $\xi$ are positive constants. The energy density $\rho_{mcg}$
can be calculated by using energy conservation equation as follows
\begin{equation}\label{10n}
\rho_{mcg}=\bigg(C_{3}+\frac{C_{4}}{a^{3(1+\alpha)(1+\zeta)}}\bigg)^\frac{1}{1+\alpha},
\end{equation}
here $C_{4}$ is an integration constant and
$C_{3}=\frac{\xi}{1+\zeta}$. In view of MCG, Friedmann equation
takes the following form
\begin{equation}\label{11n}
H^2=\frac{1}{3M^2_p}\bigg[(C_{3}+\rho_\phi^{(1+\alpha)
(1+\zeta)})^\frac{1}{1+\alpha}+\rho_{\gamma}\bigg].
\end{equation}
Under certain conditions as mentioned in GCG case, the above
Friedmann equation reduces to
\begin{equation}\label{13n}
H^2=\frac{1}{3M^2_p}(C_{3}+\rho^{(1+\alpha)(1+\zeta)}_{\phi})^\frac{1}{1+\alpha}
\sim\frac{1}{3M^2_p}(C_{3}+V^{(1+\alpha)(1+\zeta)})^\frac{1}{1+\alpha}.
\end{equation}
Next, we extract the inflationary parameters for both cases of
dissipative coefficient.

\subsection{Weak Dissipative Regime}

Here, the temperature remains the same as in case of GCG while slow
roll parameters turn out to be
\begin{eqnarray}\nonumber
\epsilon&=&\frac{M^2_P(1+\zeta)V^{(1+\alpha)(1+\zeta)-1}V'^2}{2(C_{3}+V^{(1+\alpha)(1+\zeta)})^\frac{2+\alpha}{1+\alpha}},\quad
\eta=\frac{M^2_p}{(C_{3}+V^{(1+\alpha)(1+\zeta)})^\frac{1}{1+\alpha}}\bigg(2V''\\\nonumber&+&\frac{V'^2((1+\alpha)
(1+\zeta)-1)}{V}-\frac{V^{(1+\alpha)(1+\zeta)-1}V'^2(1+\alpha)(1+\zeta)}{(C_{3}+V^{(1+\alpha)(1+\zeta)})}\bigg),\\\nonumber
\beta &=&M_p^2\bigg((C_{3}+V^{(1+\alpha )(1+\zeta )})2(2n V''-\text
n V'(n-1)\phi ^{-1}-V'\\\nonumber&\times&(4-n)(n-1)\phi
^{-1})(({2(4-n)(C_{3}+V^{(1+\lambda )(1+\zeta )})^{\frac{2+\alpha
}{1+\alpha }}})^{-1}\\\nonumber&-&\frac{3n(1+\zeta )V'^2V^{(1+\alpha
)(1+\zeta )-1}}{2(4-n)(C_{3}+V^{(1+\alpha )(1+\zeta
)})^{\frac{2+\alpha }{1+\alpha }}}\bigg).
\end{eqnarray}
Similarly, by using Eq.(\ref{6}), the number of e-folds leads to
\begin{equation}
N=\frac{1}{M^2_P}\int_{\phi_{end}}^{\phi_{*}}\frac{(C_{3}+V^{(1+\alpha)(1+\zeta)})^\frac{1}{1+\alpha}}{V'}d\phi.
\end{equation}
With the help of Eq.(\ref{7})-(\ref{9}), power spectrum, scalar
spectral index and tensor-to-scalar ratio in the form of potential
given by
\begin{eqnarray}\nonumber
\mathcal{P}_{\mathcal{R}}&=&\left(\frac{81\pi
}{4}\right)^{\frac{1}{2}}\left(\frac{M_p^{\frac{6n-15}{4-n}}\left(C_{3}+V^{(1+\alpha
)(1+\zeta )}\right)^{\frac{15-6n}{2(4-n)(1+\alpha
)}}c^{\frac{3}{4-n}}}{3^{\frac{15-6n}{2(4-n)}}C_{*}^{\frac{n+2}
{2(n+4)}}V'^{\frac{6-3n}{4-n}}6^{\frac{n+2}{4-n}}\phi
^{\frac{(n-1)(4-n)+(n-1)(n+2)}{2(4-n)}}}\right), \\\nonumber
n_s-1&=&\frac{3M_p^2}{2(C_{3}+V^{(1+\alpha )(1+\zeta
)})^{\frac{1}{1+\alpha }}}\bigg(\frac{V'^2((1+\alpha )(1+\zeta
)-1)}{V}\\\nonumber&-&\frac{V^{(1+\alpha )(1+\zeta
)-1}V'^2}{(C_{3}+V^{(1+\alpha )(1+\zeta )})}
\bigg(\frac{3}{4}(1+\zeta )+(1+\alpha )(1+\zeta
)\bigg)\\\nonumber&+&2V''-\frac{3}{2}\bigg((C_{3}+V^{(1+\alpha
)(1+\zeta )})2(2\text n V''-\text n V'(n\\\nonumber&-&1)\phi
^{-1}-V'(4-n)(n-1)\phi ^{-1}-3n(1+\zeta
)\\\label{25}&\times&V^{(1+\alpha )(1+\zeta
)-1}V'^2))(2(4-n)(C_{3}+V^{(1+\alpha )(1+\zeta
))}))^{-1}\bigg)\bigg),\\\label{14} r&=&\frac{32\text G
V'^{\frac{6-3n}{4-n}}\phi
^{\frac{(n-1)(4-n)+(n-1)(n+2)}{2(4-n)}}6^{\frac{n+2}{4-n}}C_{*}^{\frac{n+2}{2(4-n)}}
3^{\frac{7-4n}{2(4-n)}}M_p^{\frac{7-4n}{4-n}}}{9c^{\frac{3}{4-n}}\pi
^{\frac{3}{2}}\left(C_{3}+V^{(1+\alpha )(1+\zeta
)}\right)^{\frac{7-4n}{(4-n)(2+2\alpha )}}}.
\end{eqnarray}
\underline{For positive quadratic and quartic potential}, we get the
following expressions of $r$ and $n_{s}$
\begin{eqnarray}\nonumber
r&=&\frac{32G\bigg(\sigma^2\phi +\lambda_{*} \phi
^3\bigg)^{\frac{6-3n}{4-n}}\phi
^{\frac{(n-1)(4-n)+(n-1)(n+2)}{2(4-n)}}6^{\frac{n+2}{4-n}}C_{*}^{\frac{n+2}{2(4-n)}}
3^{\frac{7-4n}{2(4-n)}}}{9c^{\frac{3}{4-n}}\pi
^{\frac{3}{2}}M_p^{\frac{4n-7}{4-n}}\bigg(C_{3}+\bigg(\frac{1}{2}\sigma^2\phi
^2+\frac{\lambda_{*} }{4}\phi ^4\bigg)^{(1+\alpha )(1+\zeta
)}\bigg)^{\frac{7-4n}{(4-n)(2+2\alpha )}}},\\\nonumber
n_s-1&=&\frac{3M_p^2}{2\{C_{3}+(\frac{1}{2}\sigma^2\phi
^2+\frac{\lambda_{*} }{4}\phi ^4)^{(1+\alpha )(1+\zeta
)}\}^{\frac{1}{1+\alpha}}}\bigg[(\sigma^2\phi +\lambda_{*} \phi
^3)^2\{(1+\alpha )\\\nonumber&\times&(1+\zeta
)-1\}\bigg({\frac{1}{2}\sigma^2\phi ^2+\frac{\lambda_{*} }{4}\phi
^4}\bigg)^{-1}-\bigg(\frac{1}{2}\sigma^2\phi ^2\frac{\lambda_{*}
}{4}\phi ^4\bigg)^{(1+\alpha
)(1+\zeta)-1}\\\nonumber&\times&(\sigma^2\phi +\lambda_{*} \phi
^3)^2\bigg\{{C_{3}+\bigg(\frac{1}{2}\sigma^2\phi
^2+\frac{\lambda_{*} }{4}\phi ^4\bigg)^{(1+\alpha)(1+\zeta
)}}\bigg\}^{-1}\\\nonumber&\times&\bigg(\frac{3}{4}(1+\zeta
)+(1+\alpha)(1+\zeta )\bigg)+2(\sigma^2+3\lambda_{*} \phi
^2)-\frac{3}{2}\\\nonumber&\times&\bigg\{2\bigg(C_{3}+(\frac{1}{2}\sigma^2\phi
^2+\frac{\lambda_{*} }{4}\phi ^4)^{(1+\alpha )(1+\zeta
)}\bigg)\bigg(2n(\sigma^2+3\lambda_{*} \phi
^2)\\\nonumber&-&n(\sigma^2\phi +\lambda_{*} \phi ^3)(n-1)\phi
^{-1}-(\sigma^2\phi +\lambda_{*} \phi ^3)(4-n)(n-1)\phi
^{-1}\\\nonumber&-&3n(1+\zeta )(\frac{1}{2}\sigma^2\phi
^2+\frac{\lambda_{*} }{4}\phi ^4)^{(1+\alpha )(1+\zeta
)-1}(\sigma^2\phi +\lambda_{*} \phi
^3)^2\\\label{16}&\times&(2(4-n)(C_{3}+(\frac{1}{2}\sigma^2\phi
^2+\frac{\lambda_{*} }{4}\phi ^4)^{(1+\alpha )(1+\zeta
)}))^{-1}\bigg)\bigg\}\bigg].
\end{eqnarray}
\underline{For negative quadratic and quartic potential}, the
relations of $r$ and $n_{s}$ lead to
\begin{eqnarray}\nonumber
r&=&\frac{32G\left(-\sigma^2\phi +\lambda_{*} \phi
^3\right)^{\frac{6-3n}{4-n}}\phi
^{\frac{(n-1)(4-n)+(n-1)(n+2)}{2(4-n)}}6^{\frac{n+2}{4-n}}C_{*}^{\frac{n+2}{2(4-n)}}
3^{\frac{7-4n}{2(4-n)}}}{9c^{\frac{3}{4-n}}\pi
^{\frac{3}{2}}M_p^{\frac{4n-7}{4-n}}(C_{3}+(s-\frac{1}{2}\sigma^2\phi
^2+\frac{\lambda_{*} }{4}\phi ^4)^{(1+\alpha )(1+\zeta
)})^{\frac{7-4n}{(4-n)(2+2\alpha )}}},\\\nonumber
n_s-1&=&\frac{3M_p^2}{2\bigg\{C_{3}+\bigg(s-\frac{1}{2}\sigma^2\phi
^2+\frac{\lambda_{*} }{4}\phi ^4\bigg)^{(1+\alpha )(1+\zeta
)}\bigg\}^{\frac{1}{1+\alpha }}}\bigg[\bigg(-\sigma^2\phi
\\\nonumber&+&\lambda_{*} \phi ^3\bigg)^2\{(1+\alpha )(1+\zeta
)-1\}\bigg({s-\frac{1}{2}\sigma^2\phi ^2+\frac{\lambda_{*} }{4}\phi
^4}\bigg)^{-1}\\\nonumber&-&\frac{(s-\frac{1}{2}\sigma^2\phi
^2+\frac{\lambda_{*} }{4}\phi ^4)^{(1+\alpha )(1+\zeta
)-1}(-\sigma^2\phi +\lambda_{*} \phi
^3)^2}{C_{3}+(s-\frac{1}{2}\sigma^2\phi ^2+\frac{\lambda_{*}
}{4}\phi ^4)^{(1+\alpha )(1+\zeta
)}}\\\nonumber&\times&\bigg(\frac{3}{4}(1+\zeta )+(1+\alpha
)(1+\zeta )\bigg)+2(-\sigma^2+3\lambda_{*} \phi
^2)\\\nonumber&-&\frac{3}{2}\bigg\{\bigg(C_{3}+(s-\frac{1}{2}\sigma^2\phi
^2+\frac{\lambda_{*} }{4}\phi ^4)^{(1+\alpha )(1+\zeta
)}\bigg)2\bigg(2n(-\sigma^2\\\nonumber&+&3\lambda_{*} \phi
^2)-n(-\sigma^2\phi +\lambda_{*} \phi ^3)(n-1)\phi
^{-1}-(-\sigma^2\phi +\lambda_{*} \phi
^3)\\\nonumber&\times&(4-n)(n-1)\phi ^{-1}\bigg)-3n(1+\zeta
)(s-\frac{1}{2}\sigma^2\phi ^2+\frac{\lambda_{*} }{4}\phi
^4)^{(1+\alpha )(1+\zeta
)-1}\\\nonumber&\times&(2(4-n)(C_{3}+(s-\frac{1}{2}\sigma^2\phi
^2+\frac{\lambda_{*} }{4}\phi ^4)^{(1+\alpha )(1+\zeta
)}))^{-1}\\\label{23}&\times&(-\sigma^2\phi +\lambda_{*} \phi
^3)^2\bigg\}\bigg].
\end{eqnarray}

\subsection{Strong Dissipative Regime}

In this case, the slow roll parameters become
\begin{eqnarray}\nonumber
\epsilon&=&\frac{M^2_P(1+\zeta)V^{(1+\alpha)(1+\zeta)-1}V'^2}{2R(C_{3}+V^{(1+\alpha)(1+\zeta)})^\frac{2+\alpha}{1+\alpha}},\quad
\eta=\frac{M^2_p}{R(C_{3}+V^{(1+\alpha)(1+\zeta)})^\frac{1}{1+\alpha}}(2V''\\\nonumber&+&\frac{V'^2((1+\alpha)(1+\zeta)-1)}
{V}-\frac{V^{(1+\alpha)(1+\zeta)-1}V'^2(1+\alpha)(1+\zeta)}{(C_{3}+V^{(1+\alpha)(1+\zeta)})},\\\nonumber
\beta &=&\frac{n}{R}M_p^2\left(\frac{4V''\left(C_{3}+V^{(1+\alpha
)(1+\zeta )}\right)-(1+\zeta )V'^2V^{(1+\alpha )(1+\zeta
)-1}}{2(n+4)\left(C_{3}+V^{(1+\alpha )(1+\zeta
)}\right)^{\frac{2+\alpha }{1+\alpha }}}\right).
\end{eqnarray}
The number of e-folds turn out to be
\begin{eqnarray}\nonumber
N=\frac{1}{M^2_P}\int_{\phi_{\textmd{end}}}^{\phi_{*}}\frac{(C_{3}
+V^{(1+\alpha)(1+\zeta)})^{\frac{1}{1+\alpha}}}{V'}Rd\phi
\end{eqnarray}
The relations for perturbation, tensor-to-scalar ratio, spectral
index can be obtained by utilizing Eq.(\ref{7})-(\ref{9}) as follows
\begin{eqnarray}\nonumber
\mathcal{P}_{\mathcal{R}}&=&\left(\frac{\pi }{4}\right)^{\frac{1}{2}}
\frac{c^{\frac{9}{n+4}} \left(C_{3}+V^{(1+\alpha )(1+\zeta
)}\right)^{\frac{9}{2(1+\alpha
)(n+4)}}}{(4C_{*})^{\frac{5n+2}{2(n+4)}}\phi
^{\frac{5(n-1)(n+4)-(n-1)(5n+2)}{2(n+4)}}V'^{\frac{6-3n}{n+4}}
M_p^{\frac{9}{n+4}}3^{\frac{9}{2(n+4)}}},\\\nonumber
n_s-1&=&\frac{3(4c_{*})^{\frac{n}{n+4}}\phi
^{\frac{(n+4)(n-1)-n(n-1)}{n+4}}3^{\frac{2}{n+4}}M_p^{\frac{4}{n+4}}}{2c^{\frac{4}{n+4}}V'^{\frac{2n}{n+4}}(C_{3}+V^{(1+\alpha
)(1+\zeta )})^{\frac{2}{(1+\alpha )(n+4)}}}
\bigg[\frac{V'^2}{{V}^{1}}\bigg((1+\alpha )(1+\zeta
)\\\nonumber&\times&-1\bigg)-\frac{V^{(1+\alpha )(1+\zeta )-1}V'^{2
}}{(C_{3}+V^{(1+\alpha )(1+\zeta )})}\times
\bigg(\frac{3}{4}(1+\zeta )+(1+\alpha )(1+\zeta
)\bigg)\\\nonumber&+&2V''-\bigg(\frac{3n}{2}\bigg(4V''(C_{3}+V^{(1+\alpha
)(1+\zeta )})-(1+\zeta )V'^2V^{(1+\alpha )(1+\zeta
)-1}\bigg)\\\nonumber&\times&\bigg({2(n+4)(C_{3}+V^{(1+\alpha
)(1+\zeta )})}\bigg)^{-1}\bigg)\bigg],
\\\nonumber \text r&=&\frac{32G(4C_{*})^{\frac{5n+2}{10}}\phi
^{\frac{5(n-1)}{2}}V'^{\frac{8-5n}{5}}\left(C_{3}+V^{(1+\alpha
)(1+\zeta )}\right)^{\frac{5n-3}{20(1+\alpha
)}}}{c^{\frac{23-5n}{10}}\pi
^{\frac{3}{2}}3^{\frac{5n-3}{20}}M_p^{\frac{5n-3}{10}}}.
\end{eqnarray}

\underline{For positive quadratic and quartic potential}, the
tensor-to-scalar ratio and scalar spectral index in terms of $\phi$
are given by
\begin{eqnarray}\nonumber
r&=&\frac{32G(4C_{*})^{\frac{5n+2}{10}}\phi
^{\frac{5(n-1)}{2}}\bigg(C_{3}+(\frac{\sigma^2\phi
^2}{2}+\frac{\lambda_{*} \phi ^4}{4})^{(1+\alpha )(1+\zeta
)}\bigg)^{\frac{5n-3}{20(1+\alpha )}}}{c^{\frac{23-5n}{10}}\pi
^{\frac{3}{2}}3^{\frac{5n-3}{20}}M_p^{\frac{5n-3}{10}}\left(\sigma^2\phi
+\lambda_{*} \phi ^3\right)^{\frac{5n-8}{5}}},\\\nonumber
n_s-1&=&\frac{3c^{\frac{-4}{n+4}}(4c_{*})^{\frac{n}{n+4}}\phi
^{\frac{(n+4)(n-1)-n(n-1)}{n+4}}3^{\frac{2}{n+4}}M_p^{\frac{4}{n+4}}}{2(\sigma^2\phi
+\lambda_{*} \phi
^3)^{\frac{2n}{n+4}}(C_{3}+(\frac{1}{2}\sigma^2\phi
^2+\frac{\lambda_{*} \phi ^4}{4})^{(1+\alpha)(1+\zeta
)})^{\frac{2}{(1+\alpha )(n+4)}}}\\\nonumber&\times&
\bigg[\frac{(\sigma^2\phi +\lambda_{*} \phi ^3)^2((1+\alpha
)(1+\zeta )-1)}{\frac{1}{2}\sigma^2\phi ^2+\frac{\lambda_{*} \phi
^4}{4}}-(\sigma^2\phi +\lambda_{*} \phi ^3)^{2 }\\\nonumber&\times&
\bigg(\frac{1}{2}\sigma^2\phi ^2+\frac{\lambda_{*} \phi
^4}{4}\bigg)^{(1+\alpha )(1+\zeta )-1}
\bigg((\frac{1}{2}\sigma^2\phi ^2+\frac{\lambda_{*} \phi
^4}{4})^{(1+\alpha )(1+\zeta )}\\\nonumber&+&C_{3}\bigg)^{-1}
\bigg(\frac{3}{4}(1+\zeta )+(1+\alpha )(1+\zeta
)\bigg)+2(\sigma^2+3\lambda_{*} \phi
^2)\\\nonumber&-&\frac{3n}{2}\bigg(4(\sigma^2+3\lambda_{*} \phi
^2)(C_{3}+(\frac{1}{2}\sigma^2\phi ^2+\frac{\lambda_{*} \phi
^4}{4})^{(1+\alpha )(1+\zeta )})\\\nonumber&-&(1+\zeta)(\sigma^2\phi
+\lambda_{*} \phi ^3)^2(\frac{1}{2}\sigma^2\phi ^2+\frac{\lambda_{*}
\phi ^4}{4})^{(1+\alpha )(1+\zeta
)-1}\\\label{26}&\times&(2(n+4)(C_{3}+(\frac{1}{2}\sigma^2\phi
^2+\frac{\lambda_{*} \phi ^4}{4})^{(1+\alpha )(1+\zeta
)}))^{-1}\bigg)\bigg].
\end{eqnarray}
\underline{For negative quadratic and quartic potential}, the
tensor-to-scalar ratio and scalar spectral index in terms of $\phi$
lead to
\begin{eqnarray}\nonumber
r&=&32G(4C_{*})^{\frac{5n+2}{10}}\phi
^{\frac{5(n-1)}{2}}(-\sigma^2\phi +\lambda_{*} \phi
^3)^{\frac{8-5n}{5}}(C_{3}+(\frac{\sigma^2\phi
^2}{2}\\\nonumber&+&\frac{\lambda_{*} \phi ^4}{4})^{(1+\alpha
)(1+\zeta )})^{\frac{5n-3}{20(1+\alpha)}}(c^{\frac{23-5n}{10}}\pi
^{\frac{3}{2}}3^{\frac{5n-3}{20}}M_p^{\frac{5n-3}{10}})^{-1}.
\\\nonumber n_s-1&=&3(4c_{*})^{\frac{n}{n+4}}\phi
^{\frac{(n+4)(n-1)-n(n-1)}{n+4}}3^{\frac{2}{n+4}}M_p^{\frac{4}{n+4}}(2c^{\frac{4}{n+4}}(-\sigma^2\phi
\\\nonumber&+&\lambda_{*} \phi
^3)^{\frac{2n}{n+4}}(C_{3}+(s-\frac{1}{2}\sigma^2\phi
^2+\frac{\lambda_{*} \phi ^4}{4})^{(1+\alpha )(1+\zeta
)})^{\frac{2}{(1+\alpha )(n+4)}})^{-1}\\\nonumber&\times&
\bigg(\frac{(-\sigma^2\phi +\lambda_{*} \phi ^3)^2((1+\alpha
)(1+\zeta )-1)}{s-\frac{1}{2}\sigma^2\phi ^2+\frac{\lambda_{*} \phi
^4}{4}}-(s-\frac{1}{2}\sigma^2\phi ^2\\\nonumber&+&\frac{\lambda_{*}
\phi ^4}{4})^{(1+\alpha )(1+\zeta )-1}(-\sigma^2\phi +\lambda_{*}
\phi ^3)^{2 }((C_{3}+(s-\frac{1}{2}\sigma^2\phi
^2\\\nonumber&+&\frac{\lambda_{*} \phi ^4}{4})^{(1+\alpha )(1+\zeta
)}))^{-1}\bigg(\frac{3}{4}(1+\zeta )+(1+\alpha )(1+\zeta
)\bigg)+2\\\nonumber&\times&(-\sigma^2+3\lambda_{*} \phi
^2)-\frac{3n}{2}\bigg(4(-\sigma^2+3\lambda_{*} \phi
^2)(C_{3}+(s\\\nonumber&-&\frac{1}{2}\sigma^2\phi
^2+\frac{\lambda_{*} \phi ^4}{4})^{(1+\alpha )(1+\zeta )})-(1+\zeta
)(-\sigma^2\phi +\lambda_{*} \phi
^3)^2\\\nonumber&\times&(s-\frac{1}{2}\sigma^2\phi
^2+\frac{\lambda_{*} \phi ^4}{4})^{(1+\alpha )(1+\zeta
)-1})\\\label{21}&\times&(2(n+4)(C_{3}+(s-\frac{1}{2}\sigma^2\phi
^2+\frac{\lambda_{*} \phi ^4}{4})^{(1+\alpha )(1+\zeta
)}))^{-1}\bigg)\bigg).
\end{eqnarray}
For MCG model, the plots of $r$ and $n_{s}$ for both weak/strong
dissipative regimes for both positive/negative quadratic and quartic
potential for three different values of $n=1,-1,-2$, respectively
are shown in Figures \textbf{7-12}. The observed constraints of
scalar ratio and spectral index are displayed in the Tables
\textbf{2(a)} and \textbf{2(b)}. It is found that the results are
compatible with WMAP$9$ \cite{i31} and Planck $2015$ \cite{i32}.\\\\
\textbf{Table 2(a):} MCG Weak and Strong Dissipative Regimes with
Positive Quadratic and Quartic Potential.
\begin{table}[bht]\centering
\begin{small}
\begin{tabular}{|c|c|c|c|c|c|c|}
\hline $Sr.No$& $n$& $r(W)$ & $n_{s}(W)$&$r(S)$&$n_{s}(S)$\\
\hline$1$& $1$&$\leq0.005$&$1.00001_{-0.00001}^{+0.00001}$&$\leq0.00025$&$1.0000_{-0.00001}^{+0.0001}$\\
\hline$2$& $-1$&$\leq0.05$&$1.17_{-0.001}^{+0.001}$&$\leq0.00028$&$1.034_{-0.001}^{+0.001}$\\
\hline$3$&$-2$&$\leq0.000025$&$0.98_{-0.01}^{+0.01}$&$\leq2.5\times10^{-14}$&$0.9_{-0.1}^{+0.1}$\\
\hline
\end{tabular}
\end{small}
\end{table}\\\\

\textbf{Table 2(b):} MCG Weak and Strong Dissipative Regime with
Negative Quadratic and Quartic Potential.
\begin{table}[bht]\centering
\begin{small}
\begin{tabular}{|c|c|c|c|c|c|c|}
\hline $Sr.No$&$n$&$r(W)$&$n_{s}(W)$&$r(S)$&$n_{s}(S)$\\
\hline$1$& $1$&$\leq0.5$&$1.0009_{-0.001}^{+0.001}$&$\leq0.0052$&$1.00000_{-0.00001}^{+0.00001}$\\
\hline$2$& $-1$&$\leq0.00007$&$1.019_{-0.001}^{+0.001}$&$\leq0.00028$&$1.034_{-0.001}^{+0.001}$\\
\hline$3$&$-2$&$\leq7.2\times10^{-6}$&$1.019_{-0.001}^{+0.001}$&$\leq2.5\times10^{-8}$&$0.9_{-0.1}^{+0.1}$\\
\hline
\end{tabular}
\end{small}
\end{table}\\

\begin{figure}
\includegraphics[width=.45\linewidth]{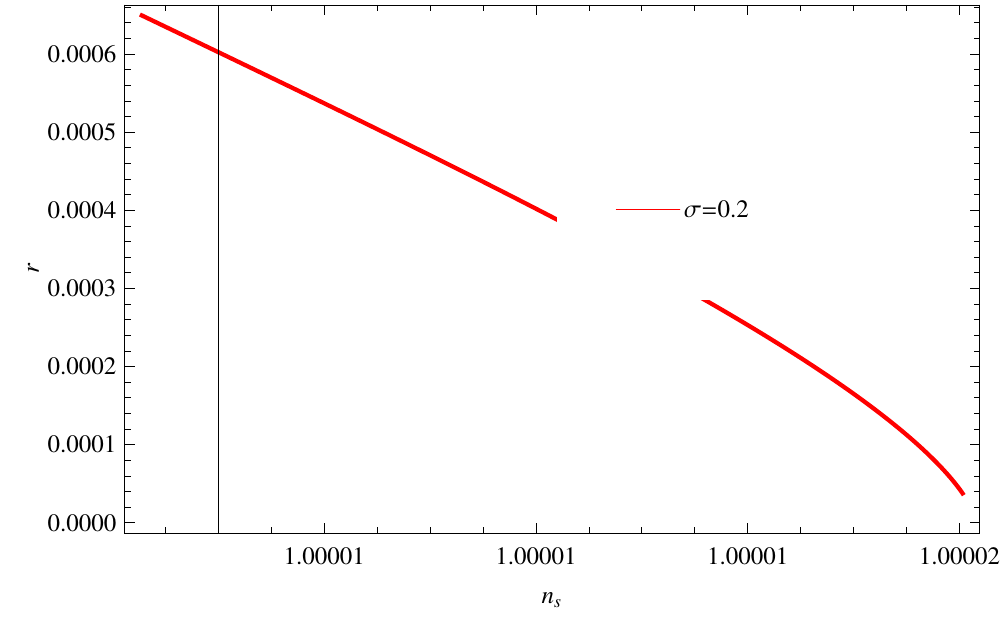}
\includegraphics[width=.45\linewidth]{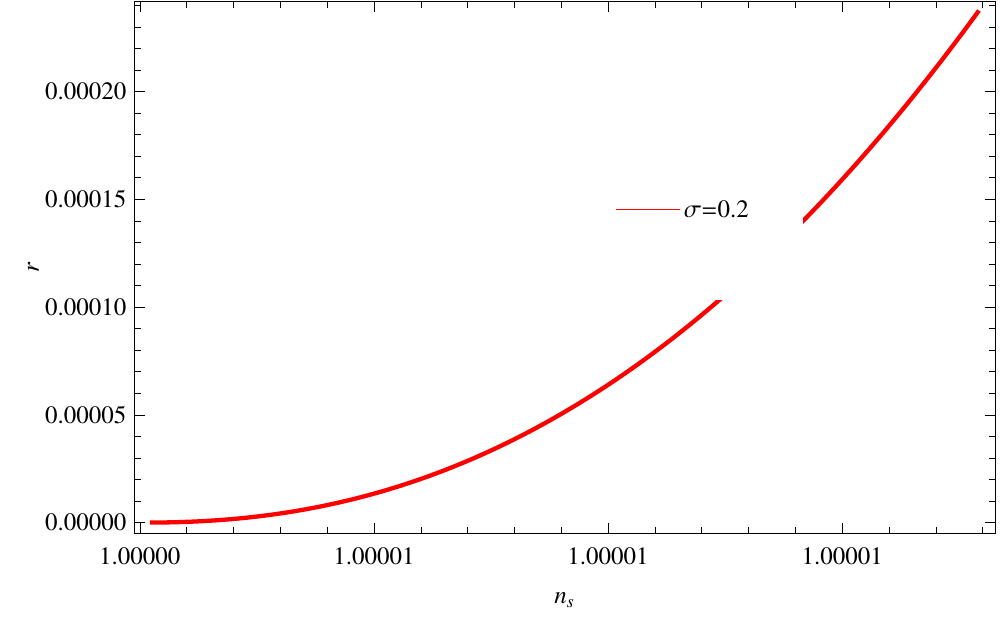}
\caption{Plot of $r$
versus $n_{s}$ for MCG model in weak (left panel) and strong (right
panel) dissipative regimes for positive potential with $n=1$.}
\end{figure}
\begin{figure}
\includegraphics[width=.45\linewidth]{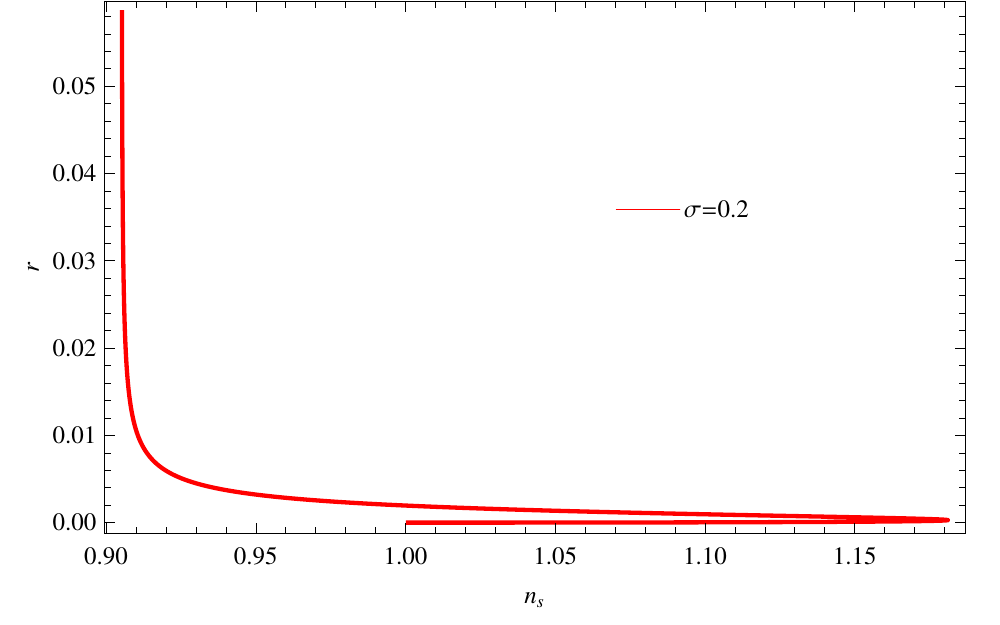}
\includegraphics[width=.45\linewidth]{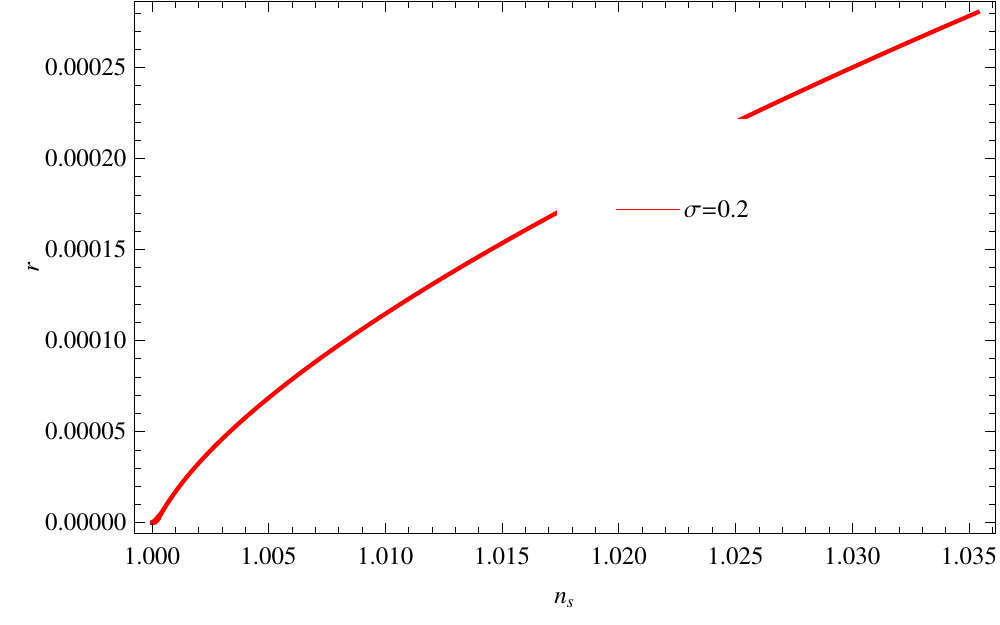}
\caption{Plot of $r$
versus $n_{s}$ for MCG model in weak (left panel) and strong (right
panel) dissipative regimes for positive potential with $n=-1$.}
\end{figure}
\begin{figure}
\includegraphics[width=.45\linewidth]{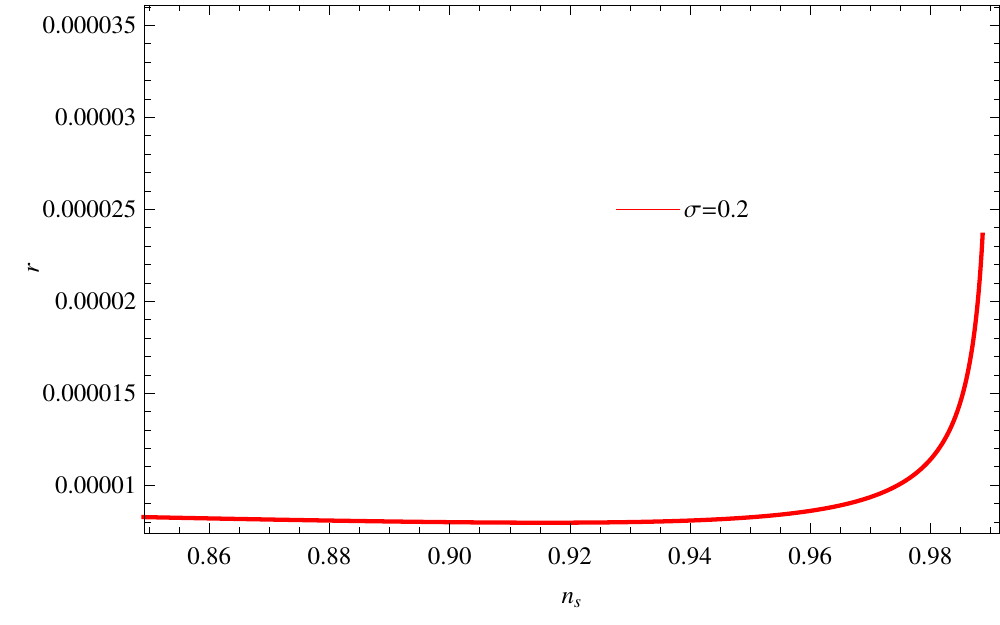}
\includegraphics[width=.45\linewidth]{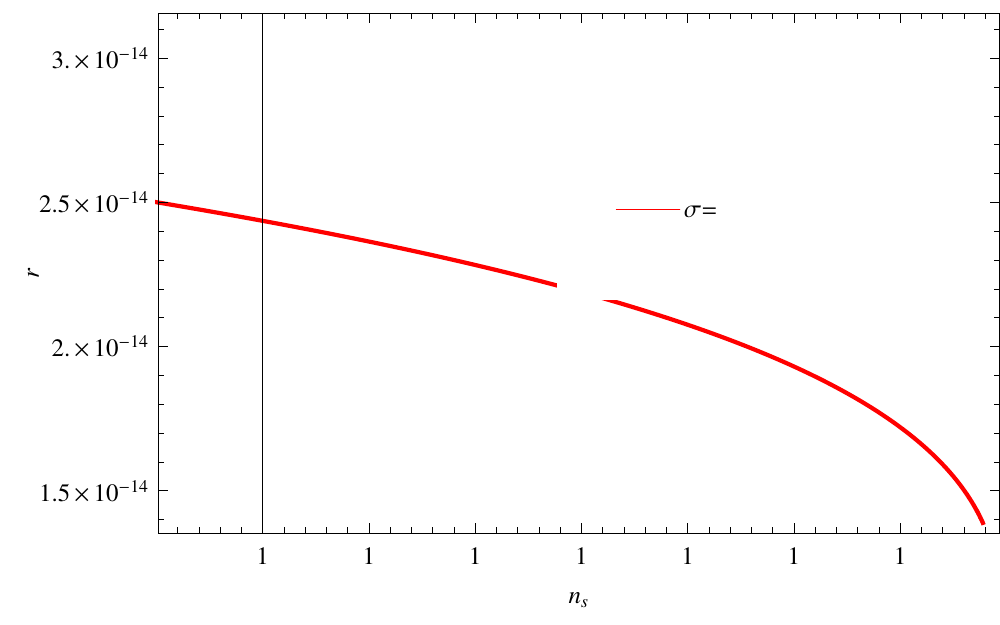}
\caption{Plot of $r$
versus $n_{s}$ for MCG model in weak (left panel) and strong (right
panel) dissipative regimes for positive potential with $n=-2$.}
\end{figure}
\begin{figure}
\includegraphics[width=.45\linewidth]{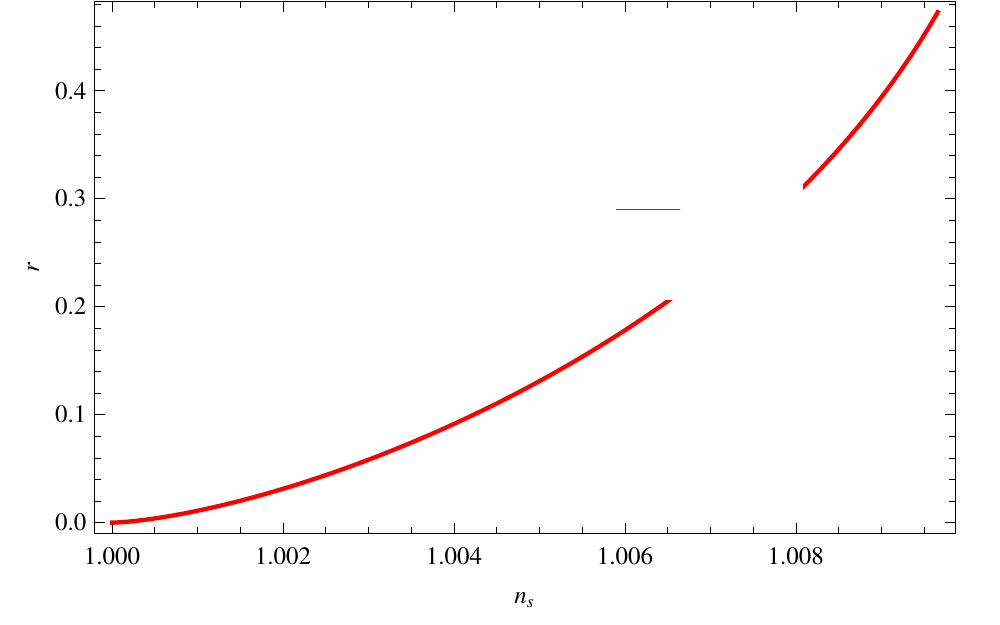}
\includegraphics[width=.45\linewidth]{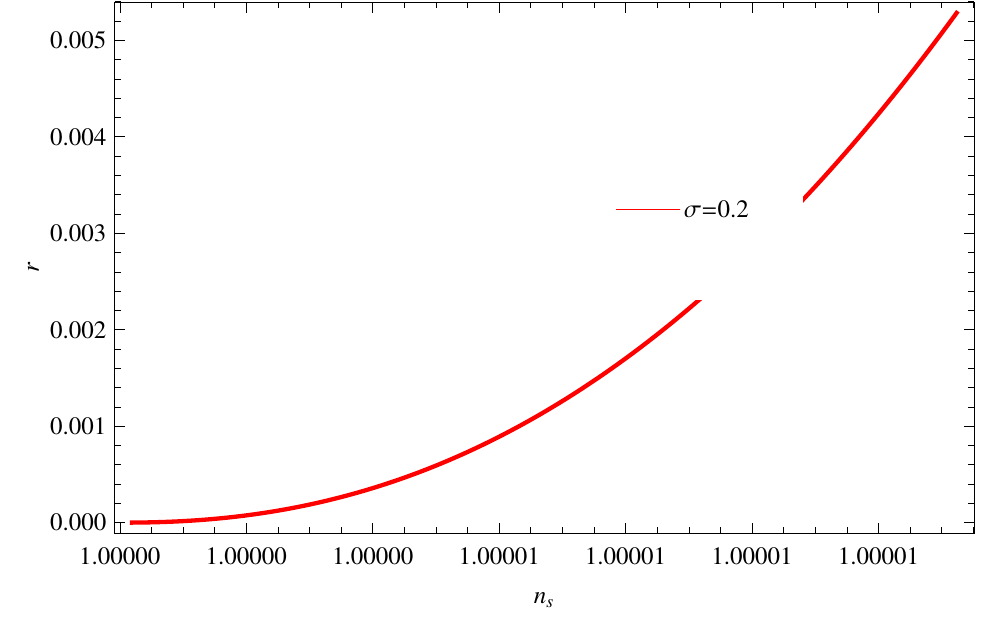}
\caption{Plot of $r$
versus $n_{s}$ for MCG model in weak (left panel) and strong (right
panel) dissipative regimes for negative potential with $n=1$.}
\end{figure}

\begin{figure}
\includegraphics[width=.45\linewidth]{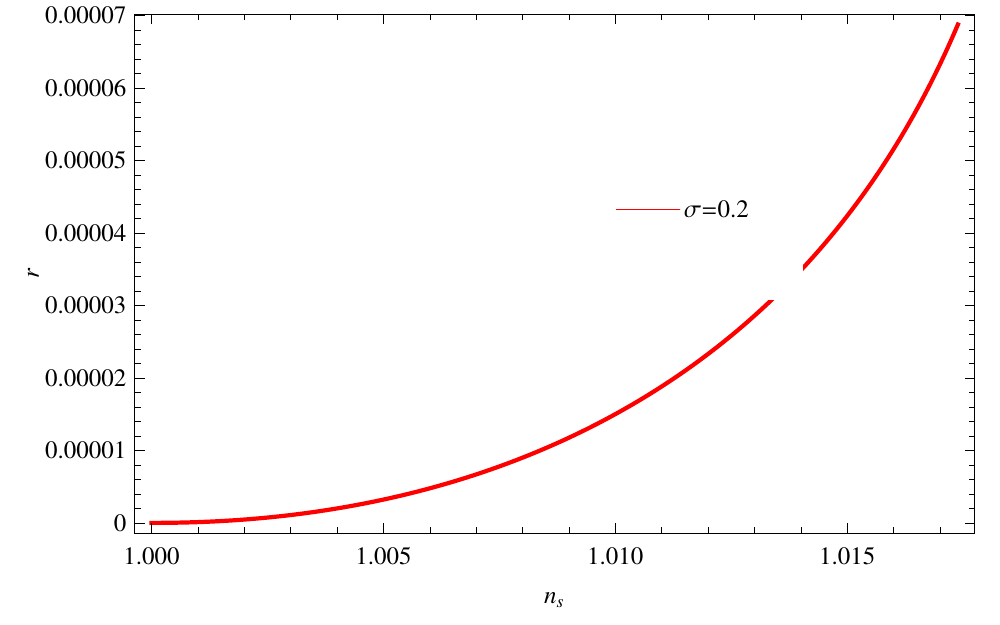}
\includegraphics[width=.45\linewidth]{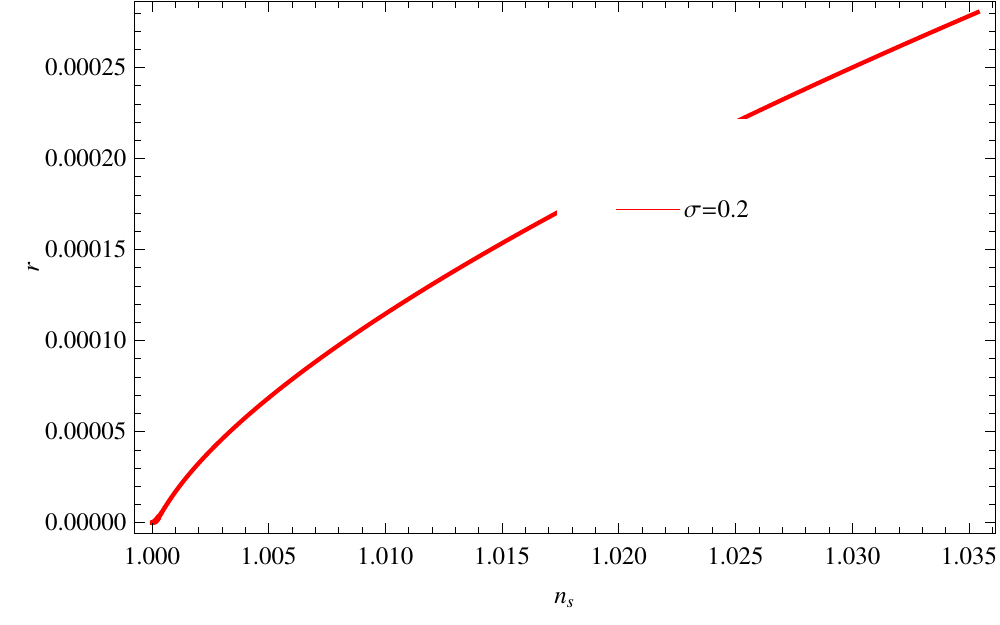}
\caption{Plot of $r$
versus $n_{s}$ for MCG model in weak (left panel) and strong (right
panel) dissipative regimes for negative potential with $n=-1$.}
\end{figure}

\begin{figure}
\includegraphics[width=.45\linewidth]{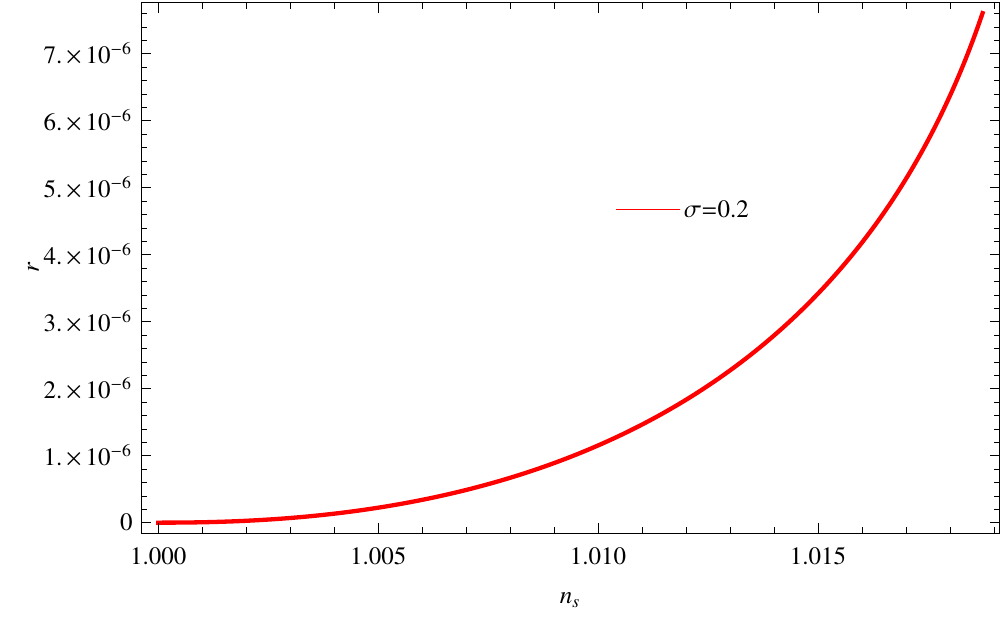}
\includegraphics[width=.45\linewidth]{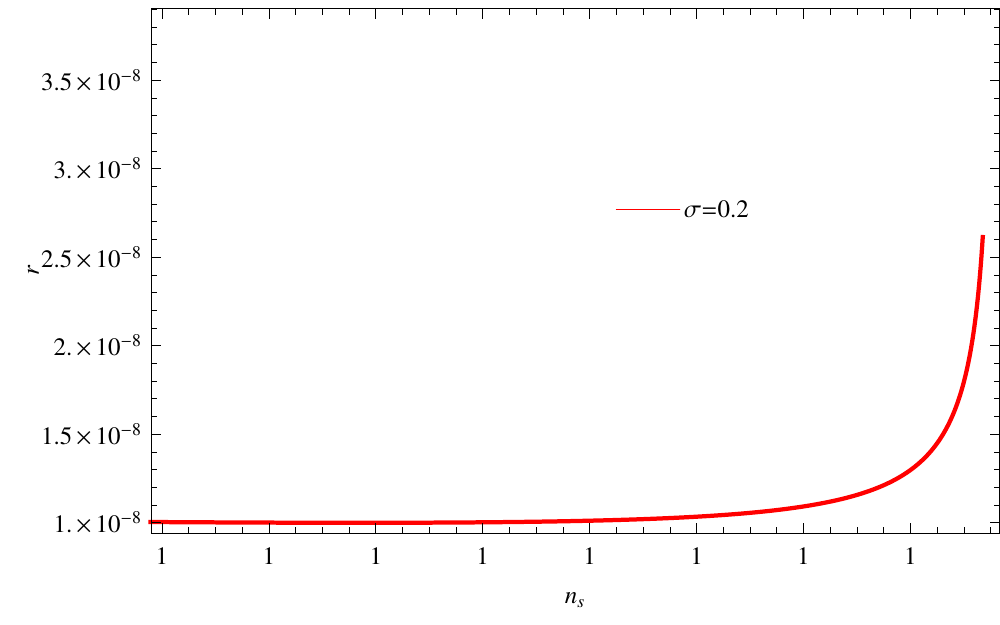}
\caption{Plot of $r$
versus $n_{s}$ for MCG model in weak (left panel) and strong (right
panel) dissipative regimes for negative potential with $n=-2$.}
\end{figure}

\section{Generalized Cosmic Chaplygin Gas}

This model is introduced by Gonzalez-Diaz \cite{18} and its EoS is
\begin{equation}\label{36}
\rho_{gccg}=-\rho^{-\alpha}\left(C_{5}+(\rho^{1+\alpha}_{gccg}-C_{5})^{-\delta}\right).
\end{equation}
Here, $C_{5}=\frac{D}{1+\delta}-1$, $D$ takes positive or negative
value, $\alpha$ is positive constant and $-l<\delta<0,~l>1$. In the
limiting case $\delta\longrightarrow0$, GCCG reduces to GCG. For
GCCG model, the energy density has the following form
\begin{equation}\label{36}
\rho_{gccg}=\left(C_{5}+(1+\frac{C_{6}}{a^{3(1+\alpha)(1+\delta)}})^\frac{1}{1+\delta}\right)^\frac{1}{1+\alpha}.
\end{equation}
The corresponding Friedmann equation becomes
\begin{equation}\label{37}
H^2=\frac{1}{3M_p^2}\left((C_{5}+(1+\rho^{(1+\alpha)(1+\delta)}_\phi)^\frac{1}{1+\delta}))^\frac{1}{1+\alpha}+\rho_{\gamma}\right).
\end{equation}
Under certain assumptions, the above Friedmann equation takes the
following form
\begin{eqnarray}\label{38}
H^2&\sim&\frac{1}{3M_p^2}\left(C_{5}+(1+V^{(1+\alpha)(1+\delta)})^\frac{1}{1+\delta}\right)^\frac{1}{1+\alpha}.
\end{eqnarray}

\subsection{Weak Dissipative Regime}

The temperature remain same as in both above cases but the slow-roll
parameters in this case take the form
\begin{eqnarray}\nonumber
\epsilon&=&\frac{M_p^2V^{(1+\alpha)(1+\delta)-1}(1+V^{(1+\alpha)(1+\delta)})^\frac{-\delta}{1+\delta}V'^2}
{2(C_{5}+(1+V^{(1+\alpha)(1+\delta)})^\frac{1}{1+\delta})^\frac{2+\alpha}{1+\alpha}},\\\nonumber
\eta&=&\frac{M^2_P}{(C_{5}+(1+V^{(1+\alpha)(1+\delta)})^\frac{1}{1+\delta})^\frac{1}{1+\alpha}}
(2V''\\\nonumber&+&\frac{V'^2((1+\alpha)(1+\delta)-1)}{V}-\frac{\delta(1+\alpha)V^{(1+\alpha)(1+\delta)-1}V'^2}
{(1+V^{(1+\alpha)(1+\delta)})}\\\nonumber&-&\frac{(1+V^{(1+\alpha)(1+\delta)})^\frac{-\delta}{1+\delta}V'^2}
{(C_{5}+(1+V^{(1+\alpha)(1+\delta)})^\frac{1}{1+\delta})}(1+\alpha)V^{(1+\alpha)(1+\delta)-1}),
\\\nonumber \beta &=& M_p^2\bigg(2(C_{5}+(1+V^{(1+\alpha )(1+\delta
)})^{\frac{1}{1+\delta }})(2\text{n V}''-\text{n V}'(n-1)\phi
^{-1}\\\nonumber&-&V'(4-n)(n-1)\phi ^{-1})-3\text{n
V}'^2V^{(1+\alpha )(1+\delta )-1}(1+V^{(1+\alpha )(1+\delta
)})^{\frac{-\delta }{1+\delta
}}\\\label{22}&\times&\bigg(2(4-n)(C_{5}+(1+V^{(1+\alpha )(1+\delta
)})^{\frac{1}{1+\delta }})^{\frac{2+\alpha }{1+\alpha
}}\bigg)^{-1}\bigg).
\end{eqnarray}
Taking into account Eq.(\ref{6}), the relation for number of e-folds
becomes
\begin{equation}\nonumber
N=\frac{1}{M^2_P}\int_{\phi_{end}}^{\phi_{*}}\frac{(C_{5}+(1+V^{(1+\alpha)(1
+\delta)})^{\frac{1}{1+\delta}})^\frac{1}{1+\alpha}}{V'}d\phi.
\end{equation}
For this model, the power spectrum, scalar spectral index and
tensor-to-scalar ratio in the form of potential can be found by
using Eq.(\ref{7})-(\ref{9}) as follow
\begin{eqnarray}
\mathcal{P}_{\mathcal{R}}&=&\left(\frac{81\pi
}{4}\right)^{\frac{1}{2}}\frac{\left(C_{5}+\left(1+V^{(1+\alpha
)(1+\delta )}\right)^{\frac{1}{1+\delta
}}\right)^{\frac{(15-6n)}{2(1+\alpha
)(4-n)}}c^{\frac{3}{4-n}}}{3^{\frac{15-6n}{2(4-n)}}M_p^{\frac{15-6n}{4-n}}
V'^{\frac{6-3n}{4-n}}6^{\frac{n+2}{4-n}}C_{*}^{\frac{n+2}{2(4-n)}}\phi
^{\frac{(n-1)(4-n)-(n-1)(n+2)}{2(4-n)}}},\\\nonumber
n_s-1&=&\frac{3M_p^2}{2(C_{5}+(1+V^{(1+\alpha)(1+\delta
)})^{\frac{1}{1+\delta }})^{1+\alpha }}\bigg(\frac{V'^{2}{((1+\alpha
)(1+\delta )-1)}}{V}\\\nonumber&-&(3\bigg(2(C_{5}+(1+V^{(1+\alpha
)(1+\delta )})^{\frac{1}{1+\delta }})(2\text{n V}''-\text{n
V}'(n-1)\phi ^{-1}\\\nonumber&-&V'(4-n)(n-1)\phi ^{-1})-3\text{n
V}'^2V^{(1+\alpha )(1+\delta )-1}(1+V^{(1+\alpha )(1+\delta
)})^{\frac{-\delta }{1+\delta
}})\\\nonumber&\times&\bigg({2(2)(4-n)(C_{5}+(1+V^{(1+\alpha
)(1+\delta )})^{\frac{1}{1+\delta }})}\bigg)^{-1}+2V''-\delta
(1+\alpha )\\\nonumber&\times&V^{(1+\alpha)(1+\delta
)-1}V'^2\bigg((1+V^{(1+\alpha)(1+\delta
)})\bigg)^{-1}-\bigg(1+V^{(1+\alpha)(1+\delta
)}\bigg)^{\frac{-\delta }{1+\delta
}}\\\nonumber&\times&V'^2(1+\alpha )V^{(1+\alpha )(1+\delta
)-1}\bigg(C_{5}+(1+V^{(1+\alpha )(1+\delta )})^{\frac{1}{1+\delta
}})\bigg)^{-1}\\\label{23}&\times&(\frac{3}{4}+(1+\alpha ))\bigg),\\
r&=&\frac{32\text{G V}'^{\frac{6-3n}{4-n}}\phi
^{\frac{(n-1)(4-n)+(n-1)(n+2)}{2(4-n)}}6^{\frac{n+2}{4-n}}c_{*}^{\frac{n+2}{2(4-n)}}
3^{\frac{7-4n}{2(4-n)}}M_p^{\frac{7-4n}{4-n}}}{9c^{\frac{3}{4-n}}\pi
^{\frac{3}{2}}\left(C_{5}+\left(1+V^{(1+\alpha )(1+\delta
)}\right)^{\frac{1}{1+\delta}}\right)^{\frac{7-4n}{2(4-n)(1+\alpha
)}}}.
\end{eqnarray}

\underline{For positive quadratic and quartic potential}, the
relations of $r$ and $n_{s}$ in terms of scalar field can be
expressed as
\begin{eqnarray}
r&=&\frac{32G\left(\sigma^2\phi +\lambda_{*} \phi
^3\right)^{\frac{6-3n}{4-n}}\phi
^{\frac{(n-1)(4-n)+(n-1)(n+2)}{2(4-n)}}6^{\frac{n+2}{4-n}}c_{*}^{\frac{n+2}
{2(4-n)}}3^{\frac{7-4n}{2(4-n)}}M_p^{\frac{7-4n}{4-n}}}{9c^{\frac{3}{4-n}}\pi
^{\frac{3}{2}}\left(C_{5}+\left(1+\left(\frac{1}{2}\sigma^2\phi
^2+\frac{\lambda_{*} }{4}\phi ^4\right)^{(1+\alpha )(1+\delta
)}\right)^{\frac{1}{1+\delta }}\right)^{\frac{7-4n}{2(4-n)(1+\alpha
)}}}, \\\nonumber
n_s-1&=&\frac{3M_p^2}{2(C_{5}+(1+(\frac{1}{2}\sigma^2\phi
^2+\frac{\lambda_{*} }{4}\phi ^4)^{(1+\alpha )(1+\delta
)})^{\frac{1}{1+\delta }})^{1+\alpha
}}\\\nonumber&\times&\bigg((\sigma^2\phi +\lambda_{*} \phi
^3)^2((1+\alpha )(1+\delta )-1)^{\text{ }}((\frac{1}{2}\sigma^2\phi
^2\\\nonumber&+&\frac{\lambda_{*} }{4}\phi
^4))^{-1}-\bigg(3\bigg(2\bigg(C_{5}+\bigg(1+(\frac{1}{2}\sigma^2\phi
^2+\frac{\lambda_{*} }{4}\phi ^4\\\nonumber&\times&)^{(1+\alpha
)(1+\delta)}\bigg)^{\frac{1}{1+\delta }}\bigg)\bigg(2n
\bigg(\sigma^2+3\lambda_{*} \phi ^2\bigg)-n(\sigma^2\phi
\\\nonumber&+&\lambda_{*} \phi ^3)(n-1)\phi ^{-1}-(\sigma^2\phi
+\lambda_{*} \phi ^3)(4-n)(n\\\nonumber&-&1)\phi
^{-1}\bigg)-3n(\sigma^2\phi +\lambda_{*} \phi
^3)^2(\frac{1}{2}\sigma^2\phi ^2+\frac{\lambda_{*} }{4}\phi
^4)^{(1+\alpha )(1+\delta
)-1}\\\nonumber&\times&\bigg(1+(\frac{1}{2}\sigma^2\phi
^2+\frac{\lambda_{*} }{4}\phi ^4)^{(1+\alpha )(1+\delta
)}\bigg)^{\frac{-\delta }{1+\delta
}}\bigg)\bigg)/\bigg(2(2)(4-n)\\\nonumber&\times&\bigg(C_{5}+\bigg(1+(\frac{1}{2}\sigma^2\phi
^2+\frac{\lambda_{*} }{4}\phi ^4)^{(1+\alpha )(1+\delta
)}\bigg)^{\frac{1}{1+\delta
}}\bigg)\bigg)\\\nonumber&+&2(\sigma^2+3\lambda_{*} \phi ^2)-\delta
(1+\alpha )(\frac{1}{2}\sigma^2\phi ^2+\frac{\lambda_{*} }{4}\phi
^4)^{(1+\alpha )(1+\delta )-1}\\\nonumber&\times&(\sigma^2\phi
+\lambda_{*} \phi ^3)^2(((1+(\frac{1}{2}\sigma^2\phi
^2+\frac{\lambda_{*} }{4}\phi ^4)^{(1+\alpha )(1+\delta
)}\bigg))^{-1}\\\nonumber&-&(1+(\frac{1}{2}\sigma^2\phi
^2+\frac{\lambda_{*} }{4}\phi ^4)^{(1+\alpha )(1+\delta
)})^{\frac{-\delta }{1+\delta }}(\sigma^2\phi +\lambda_{*} \phi
^3)^2\\\nonumber&\times&(1+\alpha )(\frac{1}{2}\sigma^2\phi
^2+\frac{\lambda_{*} }{4}\phi ^4)^{(1+\alpha )(1+\delta
)-1}\bigg(C_{5}+(1\\\label{24}&+&(\frac{1}{2}\sigma^2\phi
^2+\frac{\lambda_{*} }{4}\phi ^4)^{(1+\alpha )(1+\delta
)})^{\frac{1}{1+\delta }})\bigg)^{-1}(\frac{3}{4}+(1+\alpha
))\bigg).
\end{eqnarray}

\underline{For negative quadratic and quartic potential}, the
relations of $r$ and $n_{s}$ in terms of scalar field are given by
\begin{eqnarray}
r&=&\frac{32G\left(-\sigma^2\phi +\lambda_{*} \phi
^3\right)^{\frac{6-3n}{4-n}}\phi
^{\frac{(n-1)(4-n)+(n-1)(n+2)}{2(4-n)}}6^{\frac{n+2}{4-n}}c_{*}^{\frac{n+2}{2(4-n)}}
3^{\frac{7-4n}{2(4-n)}}M_p^{\frac{7-4n}{4-n}}}{9c^{\frac{3}{4-n}}\pi
^{\frac{3}{2}}\left(C_{5}+\left(1+\left(s-\frac{1}{2}\sigma^2\phi
^2+\frac{\lambda_{*} }{4}\phi ^4\right)^{(1+\alpha )(1+\delta
)}\right)^{\frac{1}{1+\delta }}\right)^{\frac{7-4n}{2(4-n)(1+\alpha
)}}},\\\nonumber
n_s-1&=&\frac{3M_p^2}{2\bigg(C_{5}+\bigg(1+(s-\frac{1}{2}\sigma^2\phi
^2+\frac{\lambda_{*} }{4}\phi ^4)^{(1+\alpha )(1+\delta
)}\bigg)^{\frac{1}{1+\delta }}\bigg)^{1+\alpha
}}\\\nonumber&\times&\bigg(\frac{(-\sigma^2\phi +\lambda_{*} \phi
^3)^2((1+\alpha )(1+\delta )-1)^{\text{
}}}{(s-\frac{1}{2}\sigma^2\phi ^2+\frac{\lambda_{*} }{4}\phi
^4)}-\bigg(3\bigg(2\bigg(C_{5}\\\nonumber&+&\bigg(1+(s-\frac{1}{2}\sigma^2\phi
^2+\frac{\lambda_{*} }{4}\phi ^4)^{(1+\alpha )(1+\delta
)}\bigg)^{\frac{1}{1+\delta }}\bigg)\bigg(2n
(\\\nonumber&-&\sigma^2+3\lambda_{*} \phi ^2)-n(-\sigma^2\phi
+\lambda_{*} \phi ^3)(n-1)\phi ^{-1}-(\\\nonumber&-&\sigma^2\phi
+\lambda_{*} \phi ^3)(4-n)(n-1)\phi ^{-1}\bigg)-3n(-\sigma^2\phi
\\\nonumber&+&\lambda_{*} \phi ^3)^2(s-\frac{1}{2}\sigma^2\phi ^2+\frac{\lambda_{*}
}{4}\phi ^4)^{(1+\alpha )(1+\delta
)-1}\bigg(1+(s-\frac{1}{2}\\\nonumber&\times&\sigma^2\phi
^2+\frac{\lambda_{*} }{4}\phi ^4)^{(1+\alpha )(1+\delta
)}\bigg)^{\frac{-\delta }{1+\delta
}}\bigg)\bigg)/\bigg(2(2)(4-n)\bigg(C_{5}\\\nonumber&+&\bigg(1+(s-\frac{1}{2}\sigma^2\phi
^2+\frac{\lambda_{*} }{4}\phi ^4)^{(1+\alpha )(1+\delta
)}\bigg)^{\frac{1}{1+\delta }}\bigg)\bigg)+2(-\sigma^2+3\lambda_{*}
\phi ^2)\\\nonumber&-&\frac{\delta (1+\delta
)(s-\frac{1}{2}\sigma^2\phi ^2+\frac{\lambda_{*} }{4}\phi
^4)^{(1+\alpha )(1+\delta )-1}(-\sigma^2\phi +\lambda_{*} \phi
^3)^2}{\bigg(1+(s-\frac{1}{2}\sigma^2\phi ^2+\frac{\lambda_{*}
}{4}\phi ^4)^{(1+\alpha )(1+\delta
)}\bigg)}\\\nonumber&-&\bigg(1+(s-\frac{1}{2}\sigma^2\phi
^2+\frac{\lambda_{*} }{4}\phi ^4)^{(1+\alpha )(1+\delta
)}\bigg)^{\frac{-\delta }{1+\delta }}(-\sigma^2\phi +\lambda_{*}
\phi ^3)^2(1+\alpha )\\\nonumber&\times&(s-\frac{1}{2}\sigma^2\phi
^2+\frac{\lambda_{*} }{4}\phi ^4)^{(1+\alpha )(1+\delta
)-1}(\bigg(C_{5}+\bigg(1+(s-\frac{1}{2}\sigma^2\phi
^2\\\label4&+&\frac{\lambda_{*} }{4}\phi ^4)^{(1+\alpha )(1+\delta
)}\bigg)^{\frac{1}{1+\delta
}}\bigg))^{-1}\bigg(\frac{3}{4}+(1+\alpha )\bigg)\bigg).
\end{eqnarray}

\subsection{Strong Dissipative Regime}

In this case, the slow roll parameters become
\begin{eqnarray}\nonumber
\epsilon&=&\frac{M_p^2V^{(1+\alpha)(1+\delta)-1}(1+V^{(1+\alpha)(1
+\delta)})^\frac{-\delta}{1+\delta}V'^2}{2R(C_{5}+(1+V^{(1+\alpha)(1
+\delta)})^\frac{1}{1+\delta})^\frac{2+\alpha}{1+\alpha}},\quad
\\\nonumber
\eta&=&\frac{M^2_P}{R(C_{5}+(1+V^{(1+\alpha)(1+\delta)})^\frac{1}{1
+\delta})^\frac{1}{1+\delta}}(2V''+\frac{V'^2((1+\alpha)(1+\delta)
-1)}{V}\\\nonumber&-&\frac{\delta(1+\alpha)V^{(1+\alpha)(1+\delta)
-1}V'^2}{(1+V^{(1+\alpha)(1+\delta)})}-\frac{(1+V^{(1+\alpha)(1
+\delta)})^\frac{-\delta}{1+\delta}V'^2}{(C_{5}+(1+V^{(1+\alpha)
(1+\delta)})^\frac{1}{1+\delta})}\\\nonumber&\times&(1+\alpha)V^{(1+\alpha)(1+\delta)-1}),\quad\\\nonumber
\beta&=&\text{n M}_p^2\bigg(4V''(C_{5}+(1+V^{(1+\alpha )(1+\delta
)})^{\frac{1}{1+ }})-((1\\\nonumber&+&V^{(1+\alpha )(1+\delta
)})^{\frac{-\delta}{1+\delta}}V'^2V^{(1+\alpha )(1+\delta
)-1})\bigg)(2R(n+4)\\\label4 &\times&(C_{5}+\bigg(1+V^{(1+\alpha
)(1+\delta )})^{\frac{1}{1+\delta }})^{\frac{2+\alpha }{1+\alpha
}})^{-1}.
\end{eqnarray}
The expression of number of e-folds is given by
\begin{equation}\nonumber
N=\frac{1}{M^2_P}\int_{\phi_{end}}^{\phi_{*}}\frac{(C_{5}+(1+V^{(1
+\alpha)(1+\delta)})^\frac{1}{1+\delta})^\frac{1}{1+\alpha}}{V'}Rd\phi.
\end{equation}
The perturbation quantities turn out to be
\begin{eqnarray}\nonumber
\mathcal{P}_{\mathcal{R}}&=& \left(\frac{\pi
}{4}\right)^{\frac{1}{2}}\frac{\left(C_{5}+\left(1+V^{(1+\alpha
)(1+\delta )}\right)^{\underset{1+\delta
}{1}}\right)^{\frac{9}{2(1+\alpha
)(n+4)}}c^{\frac{9}{n+4}}}{3^{\frac{9}{2(n+4)}}M_p^{\frac{9}{n+4}}(4C_{*})^{\frac{5n+2}{2(n+4)}}\phi
^{\left.\frac{5(n-1)(n+4)-(n-1)(5n+2}{2(n+4)}\right)V'^{\frac{6-3n}{n+4}}}},
\\\nonumber
r&=&\frac{32\text{G V}'^{\frac{8-5n}{5}}\phi
^{\frac{5(n-1)}{2}}(4C_{*})^{\frac{5n+2}{10}}\left(C_{5}+\left(1+V^{(1+\alpha
)(1+\delta )}\right)\right)^{\frac{5n-3}{20(1+\alpha
)}}}{c^{\frac{23-5n}{10}}\pi
^{\frac{3}{2}}3^{\frac{5n-3}{20}}M_p^{\frac{5n-3}{10}}}, \\\nonumber
n_s-1&=&\frac{3M_p^{\frac{4}{n+4}}3^{\frac{2}{n+4}}(4C_{*})^{\frac{n}{n+4}}\phi
^{\frac{(n+4)(n-1)-n(n-1)}{n+4}}
}{2c^{\frac{4}{n+4}}V'^{\frac{2n}{n+4}}(C_{5}+(1+V^{(1+\alpha
)(1+\delta )})^{\frac{1}{1+\delta }})^{\frac{2}{(n+4)(1+\alpha
)}}}\\\nonumber&\times&\bigg(\frac{V'^{2((1+\alpha )(1+\delta
)-1)}}{V}-3n\bigg(4V''(C_{5}+(1+V^{(1+\alpha )(1+\delta
)})^{\frac{1}{1+\delta }})\\\nonumber&-&(1+V^{(1+\alpha)(1+\delta
)})^{\frac{-\delta }{1+\delta }}V'^2V^{(1+\alpha )(1+\delta
)-1}\bigg)(4(n+4)(C_{5}+(1\\\nonumber&+&V^{(1+\alpha )(1+\delta
)})^{\frac{1}{1+\delta }}))^{-1}+2V''-\frac{\delta (1+\alpha
)V^{(1+\alpha )(1+\delta )-1}V'^2}{(1+V^{(1+\alpha )(1+\delta
)})}\\\nonumber&-&\frac{(1+V^{(1+\alpha)(1+\delta )})^{\frac{-\delta
}{1+\delta}}V'^2(1+\alpha)V^{(1+\alpha )(1+\delta
)-1}}{(C_{5}+(1+V^{(1+\alpha )(1+\delta )})^{\frac{1}{1+\delta
}})}\bigg(\frac{3}{4}\\\label5 &+&(1+\alpha )\bigg)\bigg).
\end{eqnarray}

\underline{For positive quadratic and quartic potential}, the
tensor-to-scalar ratio and scalar spectral index in terms of $\phi$
can be expressed as follows
\begin{eqnarray}\nonumber
r&=&32G(\sigma^2\phi +\lambda_{*} \phi ^3)^{\frac{8-5n}{5}}\phi
^{\frac{5(n-1)}{2}}(4C_{*})^{\frac{5n+2}{10}}\bigg(C_{5}+\bigg(1+(\frac{1}{2}\sigma^2\phi
^2\\\nonumber&+&\frac{\lambda_{*} }{4}\phi ^4)^{(1+\alpha )(1+\delta
)}\bigg)\bigg)^{\frac{5n-3}{20(1+\alpha )}}(c^{\frac{23-5n}{10}}\pi
 ^{\frac{3}{2}}3^{\frac{5n-3}{20}}M_p^{\frac{5n-3}{10}})^{-1},
\\\nonumber
n_s-1&=&3M_p^{\frac{4}{n+4}}3^{\frac{2}{n+4}}(4C_{*})^{\frac{n}{n+4}}\phi
^{\frac{(n+4)(n-1)-n(n-1)}{n+4}}(2c^{\frac{4}{n+4}}(\sigma^2\phi
+\lambda_{*} \phi
^3\\\nonumber&\times&)^{\frac{2n}{n+4}}(C_{5}+(1+(\frac{1}{2}\sigma^2\phi
^2+\frac{\lambda_{*} }{4}\phi ^4)^{(1+\alpha )(1+\delta
)})^{\frac{1}{1+\delta }})^{\frac{2}{(n+4)(1+\alpha
)}})^{-1}\\\nonumber&\times&\bigg(\frac{(\sigma^2\phi +\lambda_{*}
\phi ^3)^2((1+\alpha )(1+\delta )-1)}{(\frac{1}{2}\sigma^2\phi
^2+\frac{\lambda_{*} }{4}\phi ^4)}-3n(4(\sigma^2+3\lambda_{*} \phi
^2)\\\nonumber&\times&(C_{5}+(1+(\frac{1}{2}\sigma^2\phi
^2+\frac{\lambda_{*} }{4}\phi ^4)^{(1+\alpha )(1+\delta
)})^{\frac{1}{1+\delta }})-(1+(\frac{1}{2}\sigma^2\phi
^2\\\nonumber&+&\frac{\lambda_{*} }{4}\phi ^4)^{(1+\alpha )(1+\delta
)})^{\frac{-\delta }{1+\delta }}(\sigma^2\phi +\lambda_{*} \phi
^3)^2(\frac{1}{2}\sigma^2\phi ^2+\frac{\lambda_{*} }{4}\phi
^4)^{(1+\alpha )(1+\delta
)-1})\\\nonumber&\times&({4(n+4)(C_{5}+(1+(\frac{1}{2}\sigma^2\phi
^2+\frac{\lambda_{*} }{4}\phi ^4)^{(1+\alpha )(1+\delta
)})^{\frac{1}{1+\delta
}})})^{-1}+2\\\nonumber&\times&(\sigma^2+3\lambda_{*} \phi
^2)-\frac{\delta (\frac{1}{2}\sigma^2\phi ^2+\frac{\lambda_{*}
}{4}\phi ^4)^{(1+\alpha )(1+\delta )-1}(\sigma^2\phi +\lambda_{*}
\phi ^3)^2}{(1+\alpha )^{-1}(1+(\frac{1}{2}\sigma^2\phi
^2+\frac{\lambda_{*} }{4}\phi ^4)^{(1+\alpha )(1+\delta
)})}\\\nonumber&-&(1+(\frac{1}{2}\sigma^2\phi ^2+\frac{\lambda_{*}
}{4}\phi ^4)^{(1+\alpha )(1+\delta )})^{\frac{-\delta
}{1+\delta}}(\sigma^2\phi +\lambda_{*} \phi
^3)^2(1+\alpha)\\\nonumber&\times&(\frac{1}{2}\sigma^2\phi
^2+\frac{\lambda_{*} }{4}\phi ^4)^{(1+\alpha)(1+\delta
)-1}((C_{5}+(1+(\frac{1}{2}\sigma^2\phi ^2+\frac{\lambda_{*}
}{4}\phi ^4\\\label{26}&\times&)^{(1+\alpha )(1+\delta
)})^{\frac{1}{1+\delta }}))^{-1}(\frac{3}{4}+(1+\alpha ))\bigg).
\end{eqnarray}
\underline{For negative quadratic and quartic potential}, the tensor
to scalar ratio and scalar spectral index in terms of $\phi$ can be
obtained as
\begin{eqnarray}\nonumber
\text r&=&32G(-\sigma^2\phi +\lambda_{*} \phi
^3)^{\frac{8-5n}{5}}\phi
^{\frac{5(n-1)}{2}}(4C_{*})^{\frac{5n+2}{10}}(C_{5}+(1+(s-\frac{1}{2}\sigma^2\phi
^2\\\nonumber&+&\frac{\lambda_{*} }{4}\phi ^4)^{(1+\alpha
)(1+\delta)}))^{\frac{5n-3}{20(1+\alpha )}}(c^{\frac{23-5n}{10}}\pi
^{\frac{3}{2}}3^{\frac{5n-3}{20}}M_p^{\frac{5n-3}{10}})^{-1},
\\\nonumber
n_s-1&=&3M_p^{\frac{4}{n+4}}3^{\frac{2}{n+4}}(4C_{*})^{\frac{n}{n+4}}\phi
^{\frac{(n+4)(n-1)-n(n-1)}{n+4}}(2c^{\frac{4}{n+4}}(-\sigma^2\phi
+\lambda_{*} \phi
^3\\\nonumber&\times&)^{\frac{2n}{n+4}}(C_{5}+(1+(s-\frac{1}{2}\sigma^2\phi
^2+\frac{\lambda_{*} }{4}\phi^4)^{(1+\alpha )(1+\delta
)})^{\frac{1}{1+\delta }})^{\frac{2}{(n+4)(1+\alpha
)}})^{-1}\\\nonumber&\times&\bigg(\frac{(-\sigma^2\phi +\lambda_{*}
\phi ^3)^2((1+\alpha)(1+\delta )-1)}{(s-\frac{1}{2}\sigma^2\phi
^2+\frac{\lambda_{*} }{4}\phi ^4)}-3n\bigg(4(-\sigma^2+3\lambda_{*}
\phi ^2)\\\nonumber&\times&(C_{5}+(1+(s-\frac{1}{2}\sigma^2\phi
^2+\frac{\lambda_{*} }{4}\phi ^4)^{(1+\alpha )(1+\delta
)})^{\frac{1}{1+\delta
}})-(1+(s-\frac{1}{2}\\\nonumber&\times&\sigma^2\phi
^2+\frac{\lambda_{*} }{4}\phi ^4)^{(1+\alpha )(1+\delta
)})^{\frac{-\delta }{1+\delta }}(-\sigma^2\phi +\lambda_{*} \phi
^3)^2(s-\frac{1}{2}\sigma^2\phi ^2+\frac{\lambda_{*} }{4}\phi
^4\\\nonumber&\times&)^{(1+\alpha)(1+\delta
)-1})(4(n+4)(C_{5}+(1+(s-\frac{1}{2}\sigma^2\phi
^2+\frac{\lambda_{*} }{4}\phi ^4\\\nonumber&\times&)^{(1+\alpha
)(1+\delta )})^{\frac{1}{1+\delta }}))^{-1}+2(-\sigma^2+3\lambda_{*}
\phi ^2)-\delta (1+\alpha )(s-\frac{1}{2}\sigma^2\phi
^2\\\nonumber&+&\frac{\lambda_{*} }{4}\phi ^4)^{(1+\alpha )(1+\delta
)-1}(-\sigma^2\phi +\lambda_{*} \phi
^3)^2((1+(s-\frac{1}{2}\sigma^2\phi ^2+\frac{\lambda_{*} }{4}\phi
^4\\\nonumber&\times&)^{(1+\alpha )(1+\delta
)}))^{-1}-(1+(s-\frac{1}{2}\sigma^2\phi ^2+\frac{\lambda_{*}
}{4}\phi ^4)^{(1+\alpha )(1+\delta )})^{\frac{-\delta }{1+\delta
}}(-\sigma^2\\\nonumber&\times&\phi +\lambda_{*} \phi
^3)^2(1+\alpha)(s-\frac{1}{2}\sigma^2\phi ^2+\frac{\lambda_{*}
}{4}\phi ^4)^{(1+\alpha )(1+\delta
)-1}((C_{5}+(1+(s\\\label{27}&-&\frac{1}{2}\sigma^2\phi
^2+\frac{\lambda_{*} }{4}\phi ^4\bigg)^{(1+\alpha )(1+\delta
)}\bigg)^{\frac{1}{1+\delta
}}\bigg))^{-1}(\frac{3}{4}+(1+\alpha))\bigg).
\end{eqnarray}

\begin{figure}
\includegraphics[width=.45\linewidth]{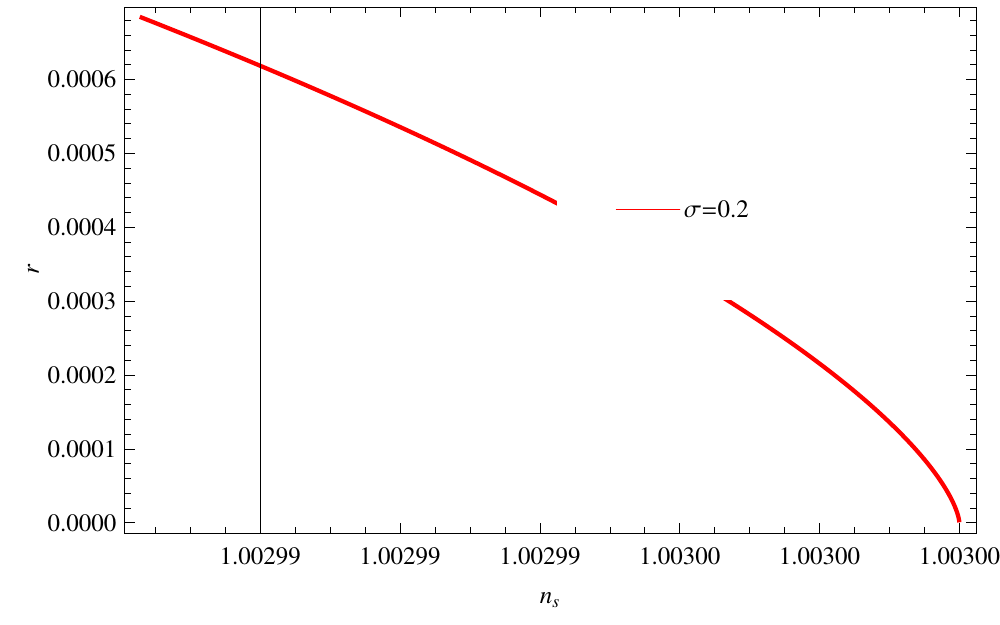}
\includegraphics[width=.45\linewidth]{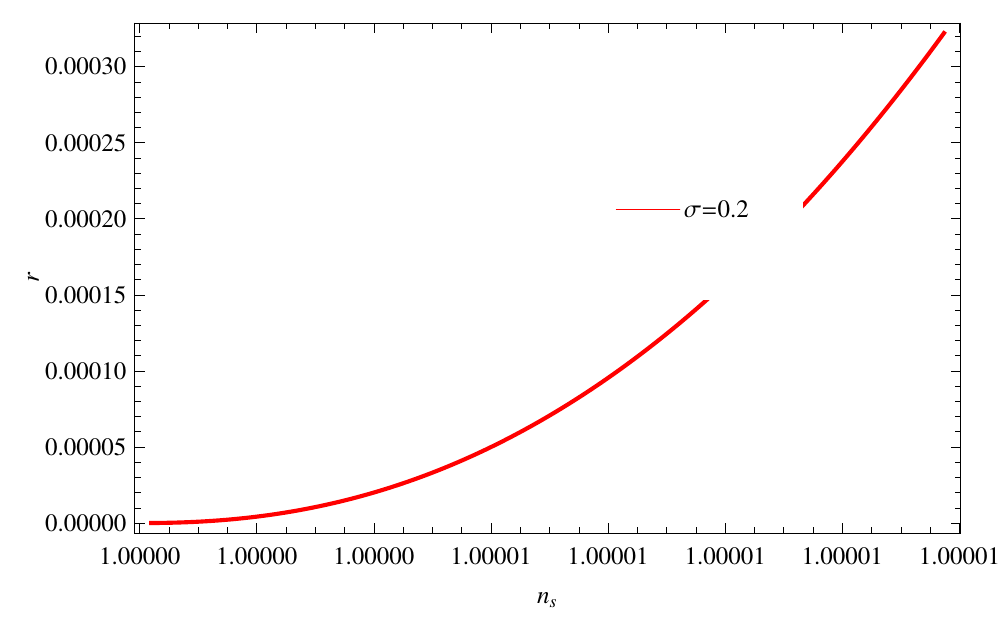}
\caption{Plot of $r$
versus $n_{s}$ for GCCG model in weak (left panel) and strong (right
panel) dissipative regimes for positive potential with $n=1$.}
\end{figure}

\begin{figure}
\includegraphics[width=.45\linewidth]{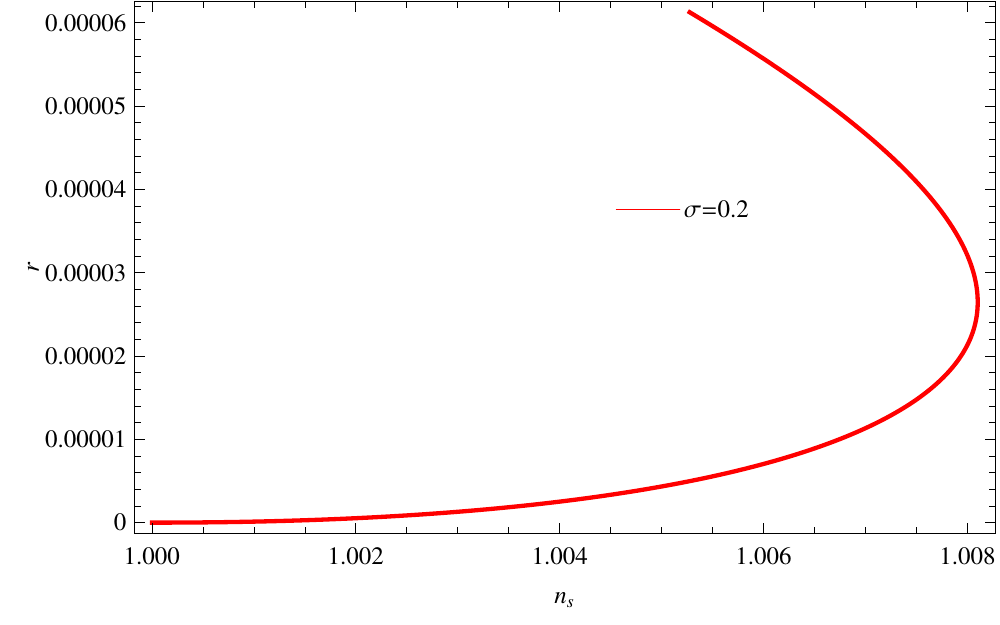}
\includegraphics[width=.45\linewidth]{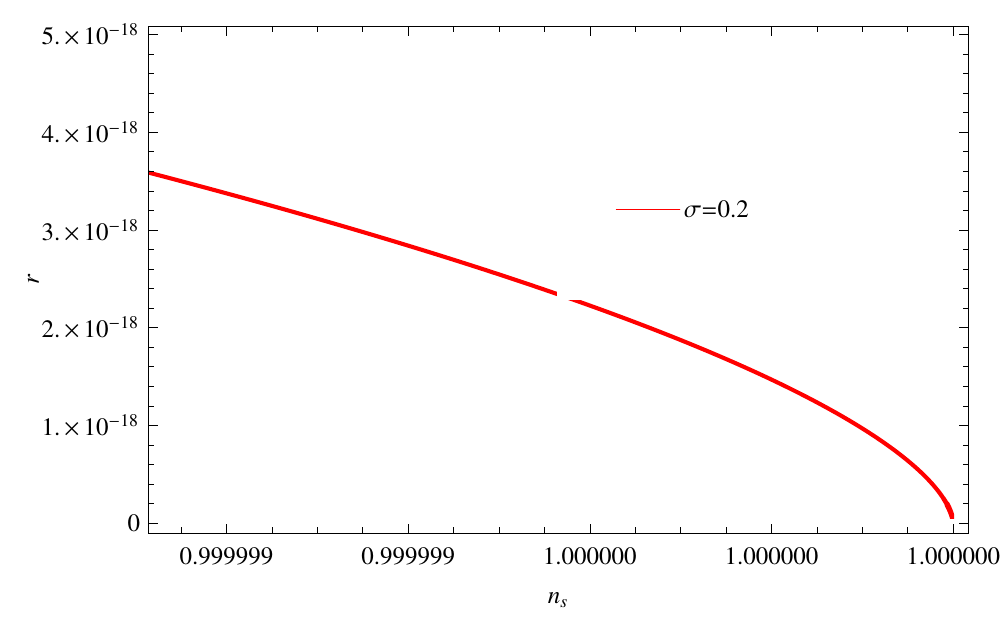}
\caption{Plot of $r$
versus $n_{s}$ for GCCG model in weak (left panel) and strong (right
panel) dissipative regimes for positive potential with $n=-1$..}
\end{figure}

\begin{figure}
\includegraphics[width=.45\linewidth]{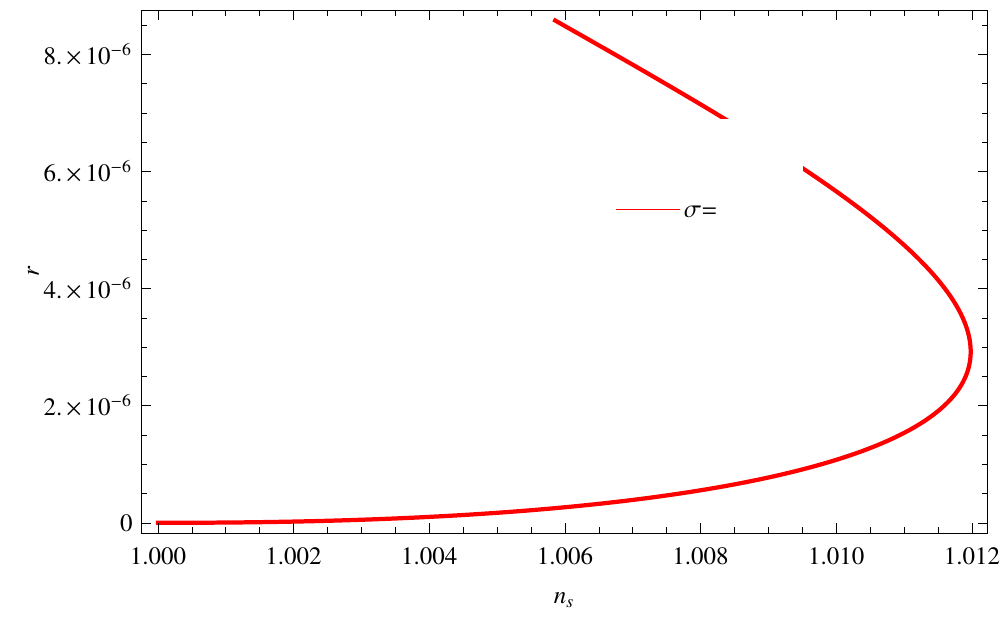}
\includegraphics[width=.45\linewidth]{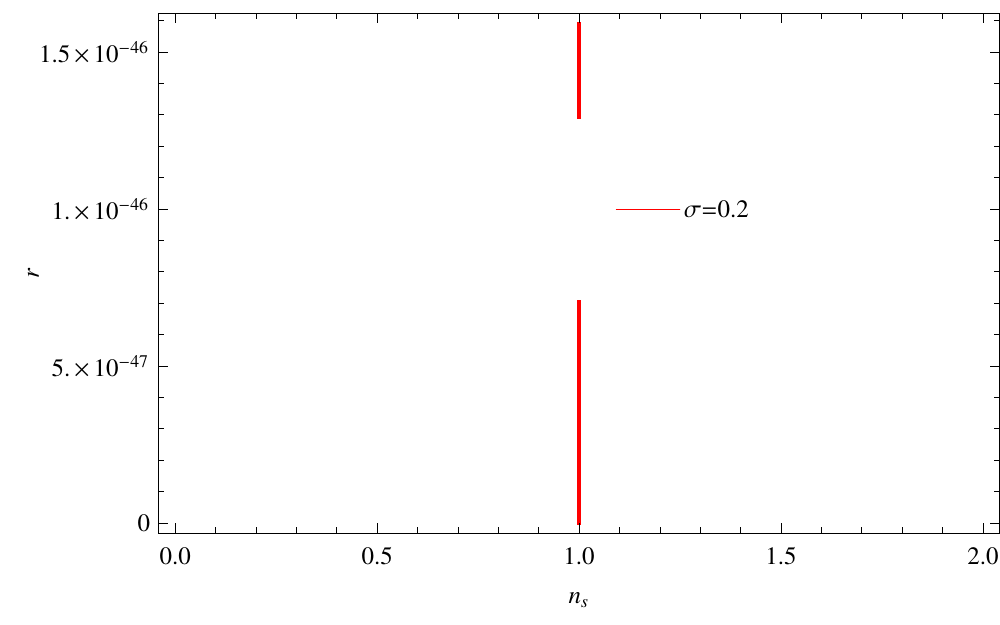}
\caption{Plot of $r$
versus $n_{s}$ for GCCG model in weak (left panel) and strong (right
panel) dissipative regimes for positive potential with $n=-2$.}
\end{figure}

\begin{figure}
\includegraphics[width=.45\linewidth]{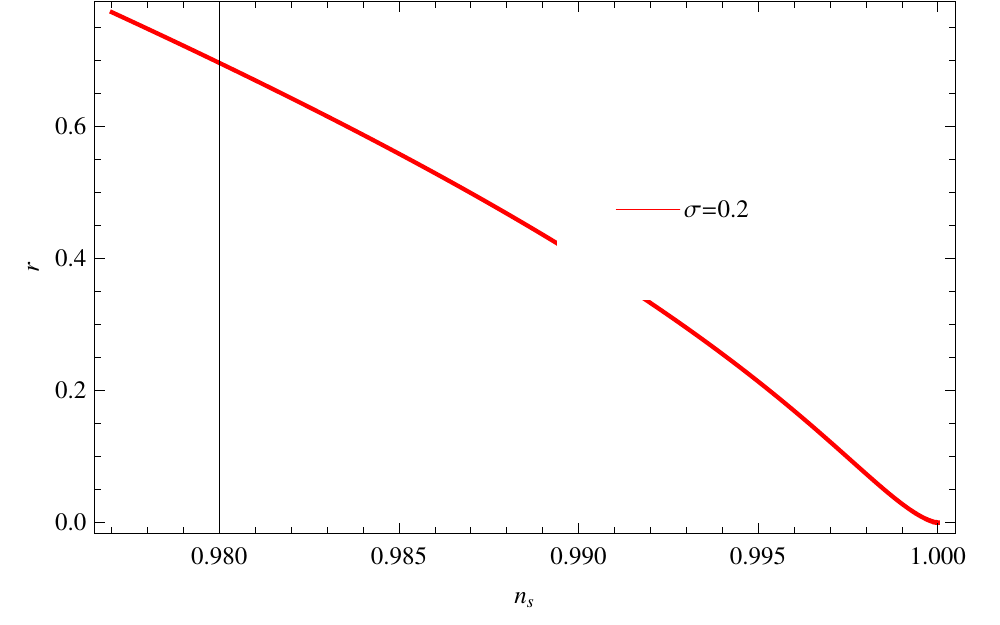}
\includegraphics[width=.45\linewidth]{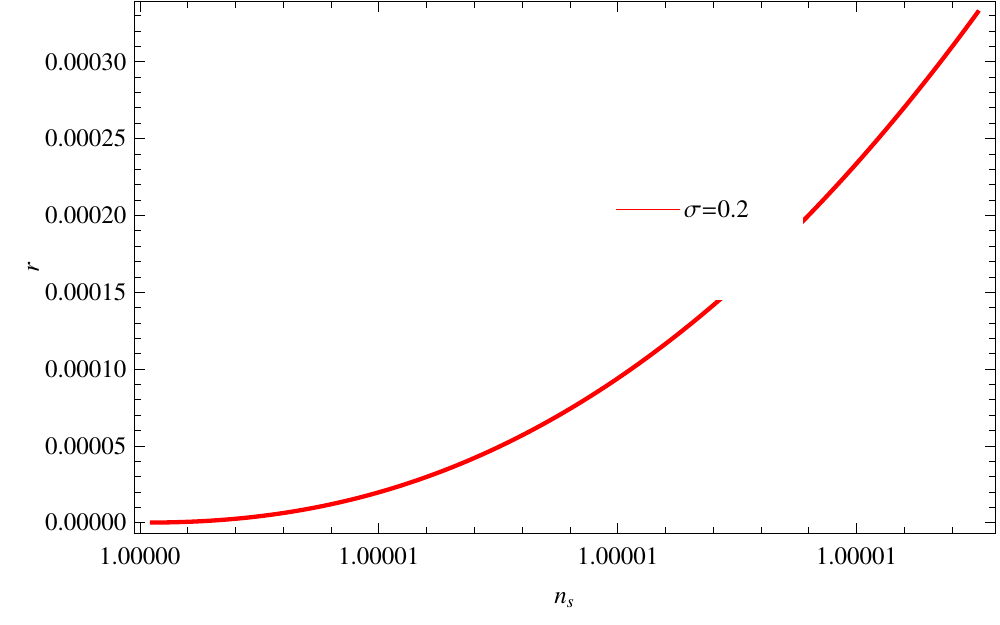}
\caption{Plot of $r$
versus $n_{s}$ for GCCG model in weak (left panel) and strong (right
panel) dissipative regimes for negative potential with $n=1$.}
\end{figure}

\begin{figure}
\includegraphics[width=.45\linewidth]{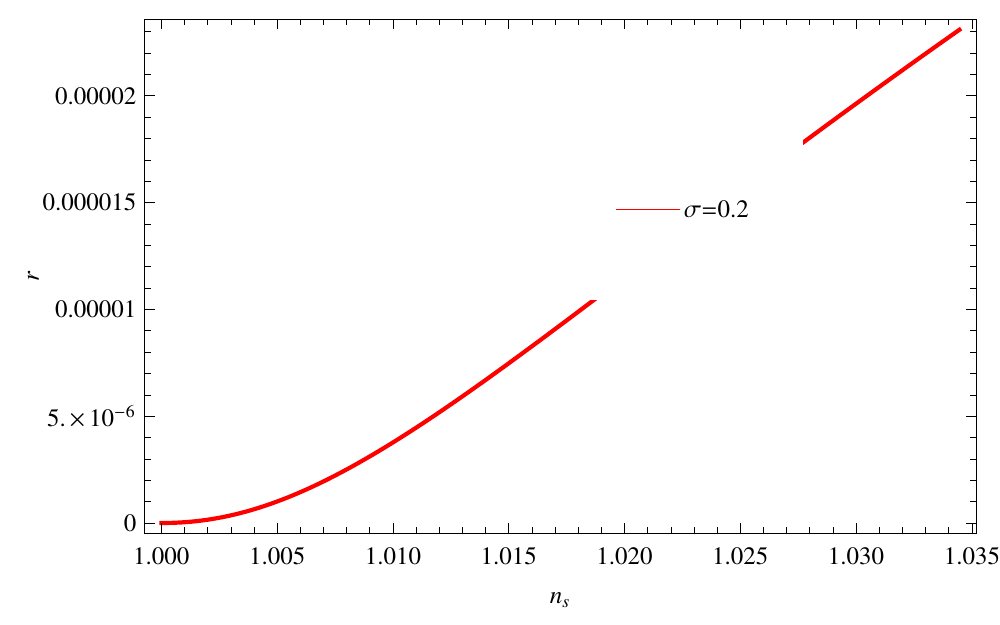}
\includegraphics[width=.45\linewidth]{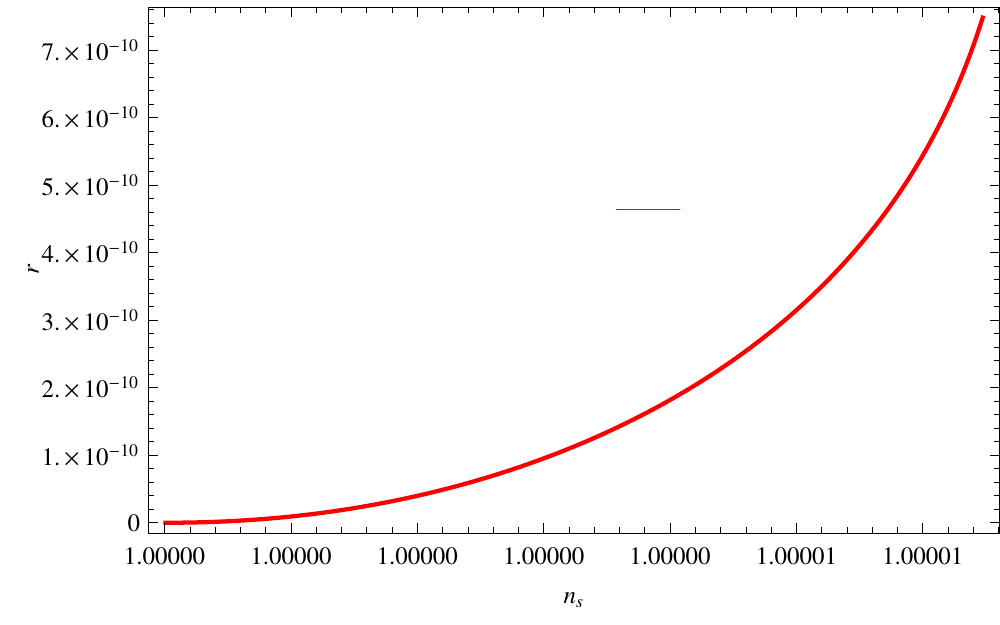}
\caption{Plot of $r$
versus $n_{s}$ for GCCG model in weak (left panel) and strong (right
panel) dissipative regimes for negative potential with $n=-1$.}
\end{figure}

\begin{figure}
\includegraphics[width=.45\linewidth]{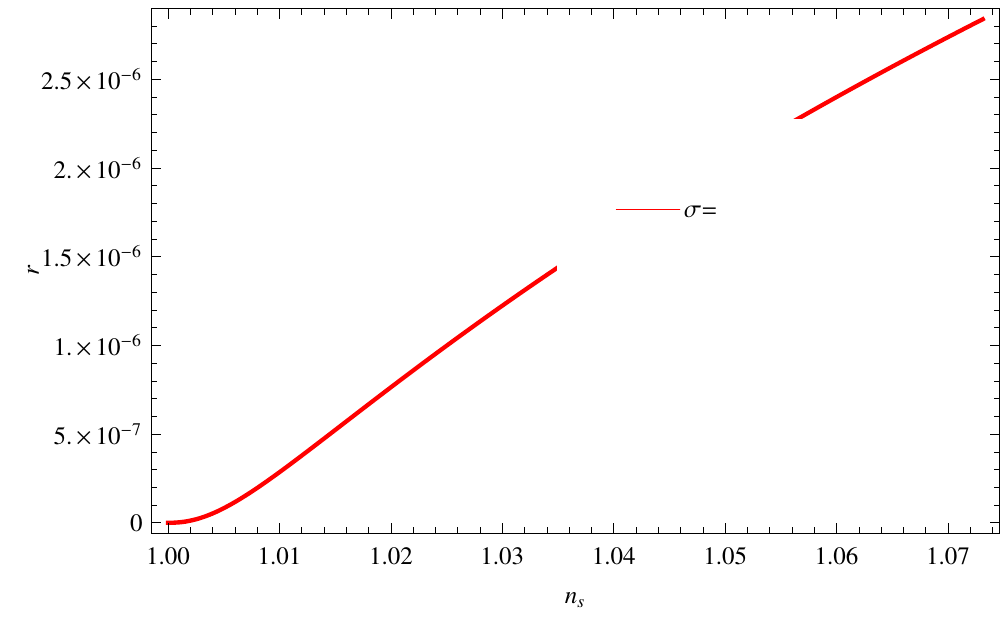}
\includegraphics[width=.45\linewidth]{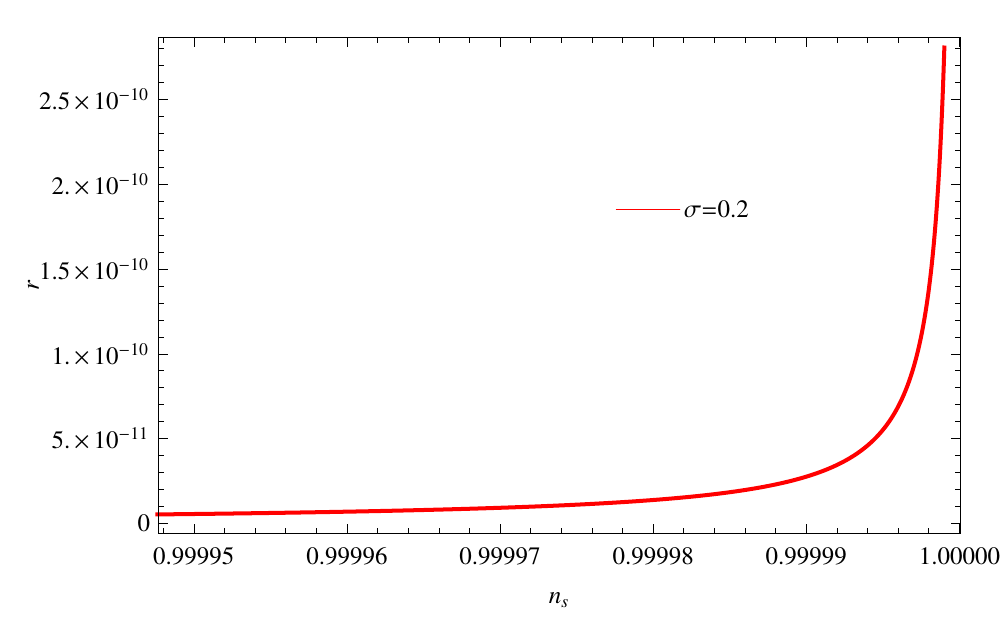}
\caption{Plot of $r$
versus $n_{s}$ for GCCG model in weak (left panel) and strong (right
panel) dissipative regimes for negative potential with $n=-2$.}
\end{figure}

For GCCG model, the plots of $r$ and $n_{s}$ for weak/strong
dissipative regimes for both positive/negative quadratic and quartic
potential for three different values of $n=1,-1,-2$, respectively
are shown in Figures \textbf{13-18}. The observed constraints of
scalar ratio and spectral index are displayed in the Tables
\textbf{3(a)} and \textbf{3(b)}. It is found that the results are
compatible with
WMAP$9$ \cite{i31} and Planck $2015$ \cite{i32}.\\
\textbf{Table 3(a):} GCCG Weak and Strong Dissipative Regime with
Positive Quadratic and Quartic Potential.
\begin{table}[bht]\centering
\begin{small}
\begin{tabular}{|c|c|c|c|c|c|c|}
\hline $Sr.No$& $n$& $r(W)$ & $n_{s}(W)$&$r(S)$&$n_{s}(S)$\\
\hline$1$& $1$&$\leq0.0007$&$1.0029_{-0.001}^{+0.001}$&$\leq0.00032$&$1.0000_{-0.0001}^{+0.0001}$\\
\hline$2$& $-1$&$\leq0.000062$&$1.004_{0.001}^{+0.001}$&$\leq3.5\times10^{-18}$&$1.000_{-0.0001}^{+0.001}$\\
\hline$3$&$-2$&$\leq8.2\times10^{-6}$&$1.011_{-0.001}^{+0.001}$&$\leq1.5\times10^{-46}$&$0.9_{-0.1}^{+0.1}$\\
\hline
\end{tabular}
\end{small}
\end{table}\\\\

\textbf{Table 3(b):} GCCG Weak and Strong Dissipative Regime with
Negative Quadratic and Quartic Potential.
\begin{table}[bht]\centering
\begin{small}
\begin{tabular}{|c|c|c|c|c|c|c|}
\hline $sr.No$& ${n}$& $r(W)$ & $n_{s}(W)$&$r(S)$&$n_{s}(S)$\\
\hline$1$& $1$&$\leq0.8$&$0.999_{-0.001}^{+0.001}$&$\leq0.00033$&$1.0000_{-0.0001}^{+0.0001}$\\
\hline$2$& $-1$&$\leq0.000025$&$1.034_{0.001}^{+0.001}$&$\leq7.2\times10^{-10}$&$1.0000_{-0.0001}^{+0.0001}$\\
\hline$3$&$-2$&$\leq2.7\times10^{-6}$&$1.06_{-0.001}^{+0.001}$&$\leq2.7\times10^{-10}$&$1.0000_{-0.0001}^{+0.0001}$\\
\hline
\end{tabular}
\end{small}
\end{table}

\section{Concluding Remarks}\label{conclusions}

We have studied warm polynomial inflation (with positive/negative
quadratic and quartic potentials) by assuming generalized form of
dissipative coefficient ($\Gamma=c\frac{T^{n}}{\phi^{n-1}}$). In
order to get the consistency of the results, we have considered
different CG models like GCG, MCG, and GCCG. We have calculated
inflationary parameters for both weak and strong dissipative regimes
such as number of e-folds, scalar spectrum, scalar spectral index
and tensor to scalar ratio. For clear analysis of results, we have
displayed the trajectories between tensor-to-scalar ratio ($r$) and
spectral index ($n_{s}$) in weak and strong dissipative regimes for
all CG models. The obtain results for all CG models in the form of
upper bounds of $r$ and $n_{s}$ are
summarized below:\\

\underline{GCG model:}

\begin{itemize}
\item For positive quadratic and quartic potential,
$r\leq0.0104168,~1.2\times 10^{-6},~1.4\times 10^{-14}$ with respect
to $n_{s}=0.98_{-0.01}^{+0.01}$, $0.96_{-0.02}^{+0.02}$,
$1.00001_{-0.00001}^{+0.00001}$, respectively, for $n=1, -1, -2$ in
weak dissipative regime. However, for strong dissipative regime,
$r\leq 0.06$, $3.0\times 10^{-15}$, $3.7\times 10^{-43}$ for
$n_{s}=0.90_{-0.06}^{+0.06}$, $0.94_{-0.02}^{+0.02}$,
$1.00011_{-0.00001}^{+0.00001}$ respectively.

\item For negative quadratic and quartic
potential, the results of tensor-to-scalar ratio and scalar spectral
index in weak dissipative regime are $r\leq 0.38$, $0.000041$,
$4.5\times 10^{-6}$ according to $n_{s}
=1.0000_{-0.0001}^{+0.0001}$, $1.0031_{-0.0001}^{+0.0001}$,
$1.0040_{-0.0001}^{+0.0001}$ respectively, and for strong
dissipative regime $r\leq0.0055$, $4.0\times 10^{-28}$, $5.2\times
10^{-44}$ for $n_{s}=1.0004_{-0.0001}^{+0.0001}$,
$1.000057_{-0.000001}^{+0.000001}$, $1.00002_{-0.00001}^{+0.00001}$
respectively.
\end{itemize}

\underline{MCG model:}

\begin{itemize}
  \item For positive quadratic and quartic potential,
$r\leq0.005$, $0.05$, $0.000025$ according to
$n_{s}=1.00001_{-0.00001}^{+0.00001}$, $1.17_{-0.001}^{+0.001}$,
$0.98_{-0.011}^{+0.01}$, respectively. Similarly, for strong
dissipative regime, the results are $r\leq0.00025$, $0.00028$,
$2.5\times 10^{-14}$ for $n_{s}=1.0000_{-0.0001}^{+0.0001}$,
$1.034_{-0.001}^{+0.001}$, $0.9_{-0.1}^{+0.1}$.
  \item The results with negative quadratic and quartic potential are $r\leq
0.5$, $0.00007$, $7.2\times 10^{-6}$ for
$n_{s}=1.0009_{-0.0001}^{+0.0001}$, $1.019_{-0.001}^{+0.001}$,
$1.019_{-0.001}^{+0.001}$ and for strong dissipative regime
$r\leq0.0052$, $0.00028$, $2.5\times 10^{-8}$ with respect to
$n_{s}=1.00000_{-0.00001}^{+0.00001}$, $1.034_{-0.001}^{+0.001}$,
$0.9_{-0.1}^{+0.1}$.
\end{itemize}

\underline{GCCG model:}
\begin{itemize}
  \item For positive quadratic and quartic potential,
$r\leq0.0007$, $0.000062$, $8.2\times 10^{-6}$ with respect to
$n_{s}=1.0029_{-0.0001}^{+0.0001}$, $1.004_{-0.001}^{+0.001}$,
$1.011_{-0.001}^{+0.001}$. In the similar way, for strong
dissipative regime, the constraints are $r\leq 0.00032$, $3.5\times
10^{-18}$, $1.5\times 10^{-46}$ for
$n_{s}=1.0000_{-0.0001}^{+0.0001}$, $1.000_{-0.001}^{+0.001}$,
$0.9_{-0.1}^{+0.1}$.
  \item For negative quadratic and quartic potential,
the results for tensor-to-scalar ratio and scalar spectral index in
weak dissipative regime are $r\leq0.8$, $0.000025$, $2.7\times
10^{-6}$ for $n_{s}=0.999_{-0.001}^{+0.001}$,
$1.034_{-0.001}^{+0.001}$, $1.06_{-0.01}^{+0.01}$, and for strong
dissipative regime, the results are $r\leq0.00033$, $7.2\times
10^{-10}$, $2.7\times 10^{-10}$ for
$n_{s}=1.0000_{-0.0001}^{+0.0001}$, $1.0000_{-0.0001}^{+0.0001}$,
$1.0000_{-0.0001}^{+0.0001}$ respectively.

\end{itemize}

It is interesting to mentioned here that the above results of $r$
and $n_s$ lie within the constraints of WMAP$9$ \cite{i31} and
Planck $2015$ \cite{i32} (as mentioned in Table \textbf{1}).

\vspace{0.5cm}

{\bf Acknowledgments}

\vspace{0.5cm}

Abdul Jawad is thankful to the Higher Education Commission,
Islamabad, Pakistan for its financial support under the grant No:
5412/Federal/NRPU/R\&D/HEC/2016 of NATIONAL RESEARCH PROGRAMME FOR
UNIVERSITIES (NRPU). N.V. was supported by Comisi\'on Nacional
de Ciencias y Tecnolog\'ia of Chile through FONDECYT Grant N$^{\textup{o}}$
3150490. Finally, the authors wish to thank the
anonymous referee for her/his valuable comments, which have helped
us to improve the presentation in our manuscript.

\end{document}